\documentclass[aps,prd,twocolumn,groupedaddress,showpacs]{revtex4}
\usepackage{graphics}
\usepackage{amssymb,amsmath,amsfonts}
\usepackage{eepic,epsfig}

\begin{document}

\title{Surface Casimir densities and induced cosmological \\
constant in higher dimensional braneworlds}
\author{Aram A. Saharian}
\email{saharyan@server.physdep.r.am}
\affiliation{Department of Physics, Yerevan State University, 1 Alex Manoogian Str.,
375025 Yerevan, Armenia \\
and Departamento de F\'{\i}sica-CCEN, Universidade Federal da Para\'{\i}ba,
58.059-970, Caixa Postal 5.008, Jo\~{a}o Pessoa, PB, Brazil}
\date{\today }

\begin{abstract}
We investigate the vacuum expectation value of the surface energy-momentum
tensor for a massive scalar field with general curvature coupling parameter
obeying the Robin boundary conditions on two codimension one parallel branes
in a $(D+1)$-dimensional background spacetime $\mathrm{AdS}_{D_{1}+1}\times
\Sigma $ with a warped internal space $\Sigma $. These vacuum densities
correspond to a gravitational source of the cosmological constant type for
both subspaces of the branes. Using the generalized zeta function technique
in combination with contour integral representations, the surface energies
on the branes are presented in the form of the sum of single brane and
second brane induced parts. For the geometry of a single brane both regions,
on the left and on the right of the brane, are considered. At the physical
point the corresponding zeta functions contain pole and finite
contributions. For an infinitely thin brane taking these regions together,
in odd spatial dimensions the pole parts cancel and the total zeta function
is finite. The renormalization procedure for the surface energies and the
structure of the corresponding counterterms are discussed. The parts in the
surface densities generated by the presence of the second brane are finite
for all nonzero values of the interbrane separation and are investigated in
various asymptotic regions of the parameters. In particular, it is shown
that for large distances between the branes the induced surface densities
give rise to an exponentially suppressed cosmological constant on the brane.
The total energy of the vacuum including the bulk and boundary contributions
is evaluated by the zeta function technique and the energy balance between
separate parts is discussed.
\end{abstract}

\pacs{04.62.+v, 03.70.+k, 11.10.Kk}
\maketitle

\section{Introduction}

\label{sec:introd}

The braneworld scenario provides an interesting alternative to the standard
Kaluza-Klein compactification of the extra dimensions. The simplest
phenomenological models describing such a scenario are the five-dimensional
Randall-Sundrum type braneworld models (for a review see \cite{Ruba01,Maar03}%
). From the point of view of embedding these models into a more fundamental
theory, such as string/M-theory, one may expect that a more complete version
of the scenario must admit the presence of additional extra dimensions
compactified on an internal manifold. From a phenomenological point of view,
the consideration of more general spacetimes offer a richer geometrical
structure and may provide interesting extensions of the Randall-Sundrum
mechanism for the geometric origin of the hierarchy. More extra dimensions
also relax the fine-tunings of the fundamental parameters. These models can
provide a framework in the context of which the stabilization of the radion
field naturally takes place. In addition, a richer topological structure of
the field configuration in transverse space provides the possibility of more
realistic spectrum of chiral fermions localized on the brane. Several
variants of the Randall-Sundrum scenario involving cosmic strings and other
global defects of various codimensions have been investigated in higher
dimensions (see, for instance, \cite{Greg00} and references therein).

Motivated by the problems of the radion stabilization and the generation of
cosmological constant, the role of quantum effects in braneworlds has
attracted great deal of attention \cite{Fabi00}-\cite{Saha06c}. A class of
higher dimensional models with the topology $\mathrm{AdS}_{D_{1}+1}\times
\Sigma $, where $\Sigma $ is a one-parameter compact manifold, and with two
branes of codimension one located at the orbifold fixed points, is
considered in Refs. \cite{Flac03b,Flac03}. In both cases of the warped and
unwarped internal manifold, the quantum effective potential induced by bulk
scalar fields is evaluated and it has been shown that this potential can
stabilize the hierarchy between the Planck and electroweak scales without
fine tuning. In addition to the effective potential, the investigation of
local physical characteristics in these models is of considerable interest.
Local quantities contain more information on the vacuum fluctuations than
the global ones and play an important role in modelling a self-consistent
dynamics involving the gravitational field. In the previous papers \cite%
{Saha06a,Saha06b} we have studied the Wightman function, the
vacuum expectation values of the field square and the
energy-momentum tensor for a scalar field with an arbitrary
curvature coupling parameter obeying Robin boundary conditions on
two codimension one parallel branes embedded in the background
spacetime $\mathrm{AdS}_{D_{1}+1}\times \Sigma $ with a warped
internal space $\Sigma $. For an arbitrary internal space $\Sigma
$, the application of the generalized Abel-Plana formula
\cite{SahaRev} allowed us to extract form the vacuum expectation
values the part due to the bulk without branes and to present the
brane induced parts in terms of exponentially convergent integrals
for the points away from the branes.

The braneworld corresponds to a manifold with boundaries and the physical
quantities, in general, receive both volume and surface contributions. In
particular, the contributions located on the visible brane are of special
interest as they are direct observables in the theory. In Ref.~\cite%
{Saha04surf} the vacuum expectation value of the surface energy-momentum
tensor is evaluated for a massive scalar field subject to Robin boundary
conditions on two parallel branes in $(D+1)$-dimensional AdS bulk (for
surface densities induced by quantum fluctuations of a bulk scalar field on
a single brane in Rindler-like spacetimes see \cite{Saha06c}). It has been
shown that for large distances between the branes the induced surface
densities give rise to an exponentially suppressed cosmological constant on
the brane. In the Randall-Sundrum braneworld model, for the interbrane
distances solving the hierarchy problem between the gravitational and
electroweak mass scales, the cosmological constant generated on the visible
brane is of the right order of magnitude with the value suggested by the
cosmological observations. In the present paper we investigate similar
issues within the framework of higher dimensional braneworld models with
warped internal spaces.

The paper is organized as follows. In the next section we describe the
corresponding bulk geometry and outline the structure of the eigenmodes and
the Kaluza-Klein spectrum. An integral representation is constructed for the
zeta function related to the vacuum expectation value of the surface
energy-momentum tensor. This function is presented in the form of the sum of
single brane and second brane induced parts. The analytic continuation of
the zeta function for the model with a single brane is done in Section \ref%
{sec:1brane}. The surface energy-momentum is investigated on both sides of
the branes and the renormalization procedure is discussed. In Section \ref%
{sec:2brane} we study the surface energy density induced by the presence of
the second brane. Various limiting cases are considered and the cosmological
constant induced on the brane is estimated. The Section \ref{sec:enbal} is
devoted to the evaluation of the total vacuum energy in the two-brane setup
on the base of the zeta function technique. The relation between the bulk
and surface energies is discussed. The last section contains a summary of
the work.

\section{Surface energy-momentum tensor and the related zeta function}

\label{sec:zetafunc}

Consider a massive scalar field $\varphi (x)$ with curvature coupling
parameter $\zeta $. The corresponding field equation has the form (we adopt
the conventions of Ref. \cite{Birrell} for the metric signature and the
curvature tensor)
\begin{equation}
\left( \nabla ^{M}\nabla _{M}+m^{2}+\zeta R\right) \varphi (x)=0,
\label{fieldeq}
\end{equation}%
where $\nabla _{M}$ is the covariant derivative operator and $R$ is the
scalar curvature for a $(D+1)$-dimensional background spacetime. For the
most important special cases of minimally and conformally coupled scalars
one has $\zeta =0$ and $\zeta =\zeta _{D}\equiv (D-1)/4D$, respectively. We
will assume that the background spacetime has the topology $\mathrm{AdS}%
_{D_{1}+1}\times \Sigma $ and is described by the line element
\begin{eqnarray}
ds^{2} &=&g_{MN}dx^{M}dx^{N}  \notag \\
&=&e^{-2k_{D}y}\left( \eta _{\mu \sigma }dx^{\mu }dx^{\sigma }-\gamma
_{ik}dX^{i}dX^{k}\right) -dy^{2},  \label{metric}
\end{eqnarray}%
where $\eta _{\mu \sigma }$ is the metric tensor for $D_{1}$-dimensional
Minkowski spacetime, $k_{D}$ is the inverse AdS radius, and the coordinates $%
X^{i}$, $i=1,\ldots ,D_{2}$, cover the internal manifold $\Sigma $, $%
D=D_{1}+D_{2}$. The Ricci scalar corresponding \ to line element (\ref%
{metric}) is given by formula $R=-D(D+1)k_{D}^{2}-e^{2k_{D}y}R_{(\gamma )}$,
with $R_{(\gamma )}$ being the scalar curvature for the metric tensor $%
\gamma _{ik}$. In the discussion below, in addition to the radial coordinate
$y$ we will also use the coordinate $z=e^{k_{D}y}/k_{D}$, in terms of which
the line element is written in the form conformally related to the metric in
the direct product spacetime $R^{(D_{1},1)}\times \Sigma $ by the conformal
factor $(k_{D}z)^{-2}$.

We are interested in one-loop vacuum effects induced by quantum fluctuations
of the bulk field $\varphi (x)$ on two parallel branes of codimension one,
located at $y=a$ and $y=b$, $a<b$. We assume that on the branes the field
obeys Robin boundary conditions
\begin{equation}
\left( \tilde{A}_{y}+\tilde{B}_{y}\partial _{y}\right) \varphi (x)=0,\quad
y=a,b,  \label{boundcond}
\end{equation}%
with constant coefficients $\tilde{A}_{y}$, $\tilde{B}_{y}$. This type of
conditions is an extension of Dirichlet and Neumann boundary conditions and
appears in a variety of situations, including the considerations of vacuum
effects for a confined charged scalar field in external fields, spinor and
gauge field theories, quantum gravity and supergravity. Robin boundary
conditions naturally arise bulk fields in braneworld models.

The branes divide the space into three regions corresponding to $-\infty
<y<a $, $a<y<b$, and $b<y<\infty $. In general, the coefficients in the
boundary conditions (\ref{boundcond}) can be different for separate regions.
In a higher dimensional generalization of the Randall-Sundrum braneworld
model the coordinate $y$ is compactified on an orbifold $S^{1}/Z_{2}$, of
ltngth $l $, $-l\leqslant y\leqslant l$, and the orbifold fixed points $y=0$
and $y=l$ are the locations of two branes. The corresponding line-element
has the form (\ref{metric}) with the warp factor $e^{-2k_{D}|y|}$. In these
models the region between the branes is employed only. For an untwisted bulk
scalar with brane mass terms $c_{a}$ and $c_{b}$, the corresponding ratio of
the coefficients in the boundary condition (\ref{boundcond}) is determined
by the expression (see, e.g., Refs. \cite{Gher00,Flac01b,Saha05} for the
case of the bulk $\mathrm{AdS}_{D+1}$ and Refs. \cite{Flac03b,Saha06a} for
the geometry under consideration)
\begin{equation}
\frac{\tilde{A}_{j}}{\tilde{B}_{j}}=-\frac{n^{(j)}c_{j}+4D\zeta k_{D}}{2}%
,\;n^{(a)}=1,\;n^{(b)}=-1.  \label{AjBjRS}
\end{equation}%
In the supersymmetric version of the model \cite{Gher00} one has $%
c_{b}=-c_{a}$ and the boundary conditions are the same for both branes. For
a twisted scalar, Dirichlet boundary conditions are obtained on both branes.

For the geometry of two parallel branes in $\mathrm{AdS}_{D_{1}+1}\times
\Sigma $ with boundary conditions (\ref{boundcond}), the Wightman function
and the vacuum expectation values (VEVs) of the field square and the bulk
energy-momentum tensor are investigated in Refs. \cite{Saha06a,Saha06b}. On
manifolds with boundaries the energy-momentum tensor in addition to the bulk
part contains a contribution located on the boundary. For an arbitrary
smooth boundary $\partial M$ with the inward-pointing unit normal vector $%
n^{L}$, the surface part of the energy-momentum tensor for a scalar field is
given by the formula \cite{Saha03emt}
\begin{equation}
T_{MN}^{\mathrm{(s)}}=\delta (x;\partial M)\tau _{MN},  \label{Ttausurf}
\end{equation}%
where the "one-sided" delta-function $\delta (x;\partial M)$ locates this
tensor on $\partial M$ and
\begin{equation}
\tau _{MN}=\zeta \varphi ^{2}K_{MN}-(2\zeta -1/2)h_{MN}\varphi n^{L}\nabla
_{L}\varphi .  \label{tausurf}
\end{equation}%
In Eq. (\ref{tausurf}), $h_{MN}=g_{MN}+n_{M}n_{N}$ is the induced metric on
the boundary and $K_{MN}=h_{M}^{L}h_{N}^{P}\nabla _{L}n_{P}$ is the
corresponding extrinsic curvature tensor.

By expanding the field operator over a complete set of eigenfunctions $%
\{\varphi _{\alpha }(x),\varphi _{\alpha }^{\ast }(x)\}$\ obeying the
boundary conditions and using the standard commutation rules, for the VEV of
the operator $\tau _{MN}$ one finds
\begin{equation}
\langle 0|\tau _{MN}|0\rangle =\sum_{\alpha }\tau _{MN}\{\varphi _{\alpha
}(x),\varphi _{\alpha }^{\ast }(x)\},  \label{modesumform}
\end{equation}%
where the bilinear form $\tau _{MN}\{\varphi ,\psi \}$ on the right is
determined by the classical energy-momentum tensor (\ref{tausurf}) and the
collective index $\alpha $ can contain both discrete and continuous parts.
Below in this section, we will consider the region between the branes. The
corresponding quantities for the regions $y\leqslant a$ and $y\geqslant b$
are obtained as limiting cases and are investigated in the next section. As
we have mentioned before, in the orbifolded version of the model, the only
bulk is the one between the branes. In this region the inward-pointing
normal to the brane at $y=j$, $j=a,b$, and the corresponding extrinsic
curvature tensor are given by the relations%
\begin{equation}
n^{(j)M}=n^{(j)}\delta _{D}^{M},\;K_{MN}^{(j)}=-n^{(j)}k_{D}h_{MN},
\label{njMKjMN}
\end{equation}%
where $n^{(j)}$ is defined in formula (\ref{AjBjRS}). By using relations (%
\ref{njMKjMN}) and the boundary conditions, the VEV of the surface
energy-momentum tensor on the brane at $y=j$ is presented in the form
\begin{equation}
\langle 0|\tau _{MN}^{(j)}|0\rangle =-h_{MN}n^{(j)}k_{D}C_{j}\langle
0|\varphi ^{2}|0\rangle _{y=j}.  \label{tauj}
\end{equation}%
with the notation%
\begin{equation}
C_{j}=\zeta -(2\zeta -1/2)\tilde{A}_{j}/(k_{D}\tilde{B}_{j}).  \label{Cjgorc}
\end{equation}%
From the point of view of physics on the brane, Eq. (\ref{tauj}) corresponds
to the gravitational source of the cosmological constant type with the
surface energy density $\varepsilon _{j}^{{\mathrm{(s)}}}=\langle 0|\tau
_{0}^{(j)0}|0\rangle $ (surface energy per unit physical volume on the brane
at $y=j$ or brane tension), stress $p_{j}^{{\mathrm{(s)}}}=-\langle 0|\tau
_{1}^{(j)1}|0\rangle $, and the equation of state
\begin{equation}
\varepsilon _{j}^{{\mathrm{(s)}}}=-p_{j}^{{\mathrm{(s)}}}.  \label{eqstate}
\end{equation}%
It is remarkable that this relation takes place for both subspaces on the
brane. It can be seen that this result is valid also for the general metric $%
g_{\mu \sigma }$ instead of $\eta _{\mu \sigma }$ in line element (\ref%
{metric}). For an untwisted bulk scalar in the higher dimensional
generalization of the Randall-Sundrum braneworld based on the bulk $%
AdS_{D_{1}+1}\times \Sigma $ the coefficient in Eq. (\ref{tauj}) is given by
the formula
\begin{equation}
C_{j}=4D\zeta (\zeta -\zeta _{D})+\frac{\zeta -1/4}{k_{D}}n^{(j)}c_{j}.
\label{gorcRS}
\end{equation}%
In particular, the corresponding surface energy vanishes for minimally and
conformally coupled scalar fields with zero brane mass terms.

In order to evaluate the expectation value (\ref{modesumform}) we need the
corresponding eigenfunctions. These functions can be taken in the decomposed
form
\begin{equation}
\varphi _{\alpha }(x^{M})=\frac{f_{n}(y)e^{-i\eta _{\mu \nu }k^{\mu }x^{\nu
}}}{\sqrt{2\omega _{\beta ,n}(2\pi )^{D_{1}-1}}}\psi _{\beta }(X),
\label{eigfunc1}
\end{equation}%
where
\begin{equation}
k^{\mu }=(\omega _{\beta ,n},\mathbf{k}),\;\omega _{\beta ,n}=\sqrt{%
k^{2}+m_{\beta ,n}^{2}},\;k=|\mathbf{k}|,  \label{omega}
\end{equation}%
and $m_{\beta ,n}$ are separation constants. The modes $\psi _{\beta }(X)$
are eigenfunctions for the internal subspace:%
\begin{equation}
\left[ \Delta _{(\gamma )}+\zeta R_{(\gamma )}\right] \psi _{\beta
}(X)=-\lambda _{\beta }^{2}\psi _{\beta }(X),  \label{psibeteq}
\end{equation}%
with the eigenvalues $\lambda _{\beta }^{2}$ and the orthonormalization
condition
\begin{equation}
\int d^{D_{2}}X\,\sqrt{\gamma }\psi _{\beta }(X)\psi _{\beta ^{\prime
}}^{\ast }(X)=\delta _{\beta \beta ^{\prime }}.  \label{normpsibet}
\end{equation}%
In Eq. (\ref{psibeteq}), $\Delta _{(\gamma )}$ is the Laplace-Beltrami
operator for the metric $\gamma _{ij}$. In the consideration below we will
assume that $\lambda _{\beta }\geqslant 0$. The dependence on $\beta $ and $%
n $ in the separation constants is factorized:%
\begin{equation}
m_{\beta ,n}^{2}=m_{n}^{2}+\lambda _{\beta }^{2},  \label{mbeta2}
\end{equation}%
where the mass spectrum $m_{n}$ is determined by the radial equation
together with the boundary conditions. From the field equation (\ref{fieldeq}%
), one obtains the equation for the function $f_{n}(y)$ with the solution
\begin{equation}
f_{n}(y)=C_{n}e^{Dk_{D}y/2}g_{\nu }^{(a)}(m_{n}z_{a},m_{n}z),  \label{fny}
\end{equation}%
where the normalization coefficient $C_{n}$ is determined below, and
\begin{equation}
g_{\nu }^{(j)}(u,v)=J_{\nu }(v)\bar{Y}_{\nu }^{(j)}(u)-Y_{\nu }(v)\bar{J}%
_{\nu }^{(j)}(u),\;j=a,b.  \label{gnuj}
\end{equation}%
Here $J_{\nu }(x)$, $Y_{\nu }(x)$ are the Bessel and Neumann functions of
the order
\begin{equation}
\nu =\sqrt{D^{2}/4-D(D+1)\zeta +m^{2}/k_{D}^{2}},  \label{nu}
\end{equation}%
and the $z$-coordinates of the branes are denoted by $z_{j}=e^{k_{D}j}/k_{D}$%
, $j=a,b$. In formula (\ref{gnuj}) and in what follows for a given function $%
F(x)$ we use the notation
\begin{equation}
\bar{F}^{(j)}(x)=A_{j}F(x)+B_{j}xF^{\prime }(x),\quad j=a,b,  \label{notbar}
\end{equation}%
with the coefficients
\begin{equation}
A_{j}=\tilde{A}_{j}+\tilde{B}_{j}k_{D}D/2,\quad B_{j}=\tilde{B}_{j}k_{D}.
\label{AjBj}
\end{equation}%
In the discussion below we will assume values of the curvature coupling
parameter for which $\nu $ is real. For imaginary $\nu $ the ground state
becomes unstable \cite{Brei82,Mezi85}. Note that for conformally and
minimally coupled massless scalar fields one has $\nu =1/2$ and $\nu =D/2$,
respectively.

The function (\ref{fny}) satisfies the boundary condition on the brane at $%
y=a$. Imposing the boundary condition on the brane $y=b$ we find that the
eigenvalues $m_{n}$ are solutions to the equation
\begin{equation}
g_{\nu }^{(ab)}(m_{n}z_{a},m_{n}z_{b})=0.  \label{cnu}
\end{equation}%
where the function on the left is defined by%
\begin{equation}
g_{\nu }^{(ab)}(u,v)=\bar{J}_{\nu }^{(a)}(u)\bar{Y}_{\nu }^{(b)}(v)-\bar{Y}%
_{\nu }^{(a)}(u)\bar{J}_{\nu }^{(b)}(v).  \label{gnab}
\end{equation}%
Equation (\ref{cnu}) determines the Kaluza-Klein (KK) spectrum along the
transverse dimension. We denote by $u=\gamma _{\nu ,n}$, $n=1,2,\ldots $,
the positive zeros of the function $g_{\nu }^{(ab)}(u,uz_{b}/z_{a})$,
arranged in the ascending order, $\gamma _{\nu ,n}<\gamma _{\nu ,n+1}$. The
eigenvalues for $m_{n}$ are related to these zeros as
\begin{equation}
m_{n}=k_{D}\gamma _{\nu ,n}e^{-k_{D}a}=\gamma _{\nu ,n}/z_{a}.
\label{mntogam}
\end{equation}%
From the orthonormality condition of the radial functions,%
\begin{equation}
\int_{a}^{b}dye^{(2-D)k_{D}y}f_{n}(y)f_{n^{\prime }}(y)=\delta _{nn^{\prime
}},  \label{normfn}
\end{equation}%
for the coefficient $C_{n}$ in Eq. (\ref{fny}) one finds
\begin{equation}
C_{n}^{2}=\frac{\pi m_{n}}{k_{D}}\frac{\bar{Y}_{\nu }^{(b)}(m_{n}z_{b})/\bar{%
Y}_{\nu }^{(a)}(m_{n}z_{a})}{\frac{\partial }{\partial u}g_{\nu
}^{(ab)}(uz_{a},uz_{b})|_{u=m_{n}}}.  \label{cn}
\end{equation}%
Note that, as we consider the quantization in the region between the branes,
$a\leqslant y\leqslant b$, the modes defined by (\ref{fny}) are normalizable
for all real values of $\nu $ from Eq. (\ref{nu}).

As it follows from formula (\ref{tauj}), the VEV of the surface
energy-momentum tensor can be obtained from the VEV of the field square
evaluated on the branes. The VEV of the field square on the bulk is derived
from the corresponding Wightman function in the coincidence limit. By using
the generalized Abel-Plana formula \cite{SahaRev} for the summation over the
zeros $\gamma _{\nu ,n}$, in Ref. \cite{Saha06a} this VEV is presented in
the form%
\begin{eqnarray}
\langle 0|\varphi ^{2}|0\rangle &=&\langle \varphi ^{2}\rangle
^{(0)}+\langle \varphi ^{2}\rangle ^{(j)}-2k_{D}^{D-1}z^{D}\beta _{D_{1}}
\notag \\
&&\times \sum_{\beta }|\psi _{\beta }(X)|^{2}\int_{\lambda _{\beta
}}^{\infty }du\,u(u^{2}-\lambda _{\beta }^{2})^{D_{1}/2-1}  \notag \\
&&\times \Omega _{j\nu }(uz_{a},uz_{b})G_{\nu }^{(j)2}(uz_{j},uz),
\label{phi2bulk}
\end{eqnarray}%
where $j=a$ and $j=b$ provide two equivalent representations and%
\begin{equation}
\beta _{D_{1}}=\frac{1}{(4\pi )^{D_{1}/2}\Gamma \left( D_{1}/2\right) }.
\label{betaD}
\end{equation}%
Here and in what follows we use the notations
\begin{eqnarray}
\Omega _{a\nu }(u,v) &=&\frac{\bar{K}_{\nu }^{(b)}(v)}{\bar{K}_{\nu
}^{(a)}(u)G_{\nu }^{(ab)}(u,v)},  \label{Omegaa} \\
\Omega _{b\nu }(u,v) &=&\frac{\bar{I}_{\nu }^{(a)}(u)}{\bar{I}_{\nu
}^{(b)}(v)G_{\nu }^{(ab)}(u,v)},  \label{Omegab}
\end{eqnarray}%
with the modified Bessel functions $I_{\nu }(u)$, $K_{\nu }(u)$, and
\begin{eqnarray}
G_{\nu }^{(j)}(u,v) &=&I_{\nu }(v)\bar{K}_{\nu }^{(j)}(u)-K_{\nu }(v)\bar{I}%
_{\nu }^{(j)}(u),  \label{Gnuj12} \\
G_{\nu }^{(ab)}(u,v) &=&\bar{K}_{\nu }^{(a)}(u)\bar{I}_{\nu }^{(b)}(v)-\bar{I%
}_{\nu }^{(a)}(u)\bar{K}_{\nu }^{(b)}(v).  \label{Gnuab}
\end{eqnarray}%
The first term on the right of Eq. (\ref{phi2bulk}), $\langle \varphi
^{2}\rangle ^{(0)}$, corresponds to the VEV in $\mathrm{AdS}_{D_{1}+1}\times
\Sigma $ spacetime without branes, $\langle \varphi ^{2}\rangle ^{(j)}$ is
induced by a single brane located at $y=j$ when the second brane is absent,
and the last term is induced by the presence of the second brane. After the
standard renormalization of the boundary free part, the VEV given by Eq. (%
\ref{phi2bulk}) is finite in the bulk and diverges on the branes. As a
result, the VEVs of the field square on the branes cannot be obtained
directly from this formula and an additional renormalization procedure is
necessary. One possibility is to remove from (\ref{phi2bulk}) the terms
which diverge for the points on the branes. An alternative way is to
restrict from the begining \ to the fluctuations on the brane and to use the
dimensional regularization or the generalized zeta function technique (for a
discussion of the relation between different methods see, for instance, Ref.
\cite{Pujo05}). Here we will follow the last method.

Substituting the eigenfunctions (\ref{eigfunc1}) into the corresponding mode
sum and integrating over the angular part of the vector $\mathbf{k}$, for
the expectation value of the energy density on the brane at $y=j$ we obtain
\begin{eqnarray}
\varepsilon _{j}^{\text{(s)}} &=&-2n^{(j)}C_{j}k_{D}^{D}z_{j}^{D}B_{j}\beta
_{D_{1}-1}\int_{0}^{\infty }dk\,k^{D_{1}-2}  \notag \\
&&\times \sum_{\beta }\sum_{n=1}^{\infty }\frac{|\psi _{\beta
}(X)|^{2}m_{n}g_{\nu }^{(l)}(m_{n}z_{l},m_{n}z_{j})}{\omega _{\beta
,n}\left. \frac{\partial }{\partial u}g_{\nu
}^{(ab)}(uz_{a},uz_{b})\right\vert _{u=m_{n}}},  \label{phi2j}
\end{eqnarray}%
where $j,l=a,b$, and $l\neq j$, $\omega _{\beta ,n}$ is defined by formulas (%
\ref{omega}), (\ref{mbeta2}), and we have used the relation $g_{\nu
}^{(j)}(u,u)=2B_{j}/\pi $. To regularize the divergent expression on the
right of this formula we define the function
\begin{eqnarray}
\Phi _{j}(s) &=&2z_{j}^{D}\frac{B_{j}\beta _{D_{1}-1}}{\mu ^{1+s}}%
\sum_{\beta }|\psi _{\beta }(X)|^{2}\int_{0}^{\infty }dk\,k^{D_{1}-2}  \notag
\\
&&\times \sum_{n=1}^{\infty }\omega _{\beta ,n}^{s}\frac{m_{n}g_{\nu
}^{(l)}(m_{n}z_{l},m_{n}z_{j})}{\frac{\partial }{\partial u}g_{\nu
}^{(ab)}(uz_{a},uz_{b})|_{u=m_{n}}},  \label{IAs}
\end{eqnarray}%
where an arbitrary mass scale $\mu $\ is introduced to keep the dimension of
the expression. After the evaluation of the integral over $k$, this
expression can be presented in the form%
\begin{equation}
\Phi _{j}(s)=\frac{z_{j}^{D}B_{j}}{(4\pi )^{(D_{1}-1)/2}}\sum_{\beta }|\psi
_{\beta }(X)|^{2}\zeta _{j\beta }(s),  \label{Fjs2}
\end{equation}%
where the generalized partial zeta function
\begin{eqnarray}
\zeta _{j\beta }(s) &=&\frac{\Gamma (-\alpha _{s})}{\Gamma (-s/2)\mu ^{s+1}}%
\sum_{n=1}^{\infty }(m_{n}^{2}+\lambda _{\beta }^{2})^{\alpha _{s}}  \notag
\\
&&\times \frac{m_{n}g_{\nu }^{(l)}(m_{n}z_{l},m_{n}z_{j})}{\frac{\partial }{%
\partial u}g_{\nu }^{(ab)}(uz_{a},uz_{b})|_{u=m_{n}}},  \label{zetsx}
\end{eqnarray}%
is introduced with the notation%
\begin{equation}
\alpha _{s}=(D_{1}+s-1)/2.  \label{alfas}
\end{equation}%
The computation of the VEV of the surface energy-momentum tensor requires
the analytic continuation of the function $\Phi _{j}(s)$ to the value $s=-1$
(here and below $|_{s=-1}$ is understood in the sense of the analytic
continuation),
\begin{equation}
\varepsilon _{j}^{\mathrm{(s)}}=-n^{(j)}k_{D}^{D}C_{j}\Phi _{j}(s)|_{s=-1}.
\label{IFs0}
\end{equation}

In order to obtain this analytic continuation we will follow the procedure
multiply used for the evaluation of the Casimir energy (see, for instance,
\cite{Bord01}). The starting point is the representation of the function (%
\ref{zetsx}) in terms of the contour integral
\begin{eqnarray}
\zeta _{j\beta }(s) &=&\frac{\Gamma (-\alpha _{s})\mu ^{-s-1}}{2\pi i\Gamma
(-s/2)}\int_{C}du\,u(u^{2}+\lambda _{\beta }^{2})^{\alpha _{s}}  \notag \\
&&\times \frac{g_{\nu }^{(l)}(uz_{l},uz_{j})}{g_{\nu }^{(ab)}(uz_{a},uz_{b})}%
,  \label{intzetsx1}
\end{eqnarray}%
where $C$ is a closed counterclockwise contour in the complex $u$ plane
enclosing all zeros $m_{n}$. The location of these zeros \ enables one to
deform the contour $C$ into a segment of the imaginary axis $(-iR,iR)$ and a
semicircle of radius $R$, $R\rightarrow \infty $, in the right half-plane.
We will also assume that the branch points $\pm i\lambda _{\beta }$ are
avoided by small semicircles $C_{\rho }^{\pm }$ in the right half-plane with
small radius $\rho $. Here we assume that there is no zero mode
corresponding to $m_{n}=0$. The changes introduced by the presence of the
zero mode will be discussed below in this section. For negative ${\mathrm{Re}%
}\,s$ with sufficiently large absolute value the integral over the large
semicircle in Eq. (\ref{intzetsx1}) tends to zero in the limit $R\rightarrow
\infty $, and the expression on the right can be transformed to%
\begin{eqnarray}
\zeta _{j\beta }(s) &=&\frac{\Gamma (-\alpha _{s})}{\Gamma \left( -\frac{s}{2%
}\right) \mu ^{s+1}}\left[ \sum_{\alpha =\pm }\int_{C_{\rho }^{\alpha }}du\,u%
\frac{(u^{2}+\lambda _{\beta }^{2})^{\alpha _{s}}}{2\pi i}\right.  \notag \\
&&\times \frac{g_{\nu }^{(l)}(uz_{l},uz_{j})}{g_{\nu }^{(ab)}(uz_{a},uz_{b})}%
-\frac{\sin \pi \alpha _{s}}{\pi }\int_{\lambda _{\beta }+\rho }^{\infty
}du\,u  \notag \\
&&\left. \times (u^{2}-\lambda _{\beta }^{2})^{\alpha _{s}}\frac{G_{\nu
}^{(l)}(uz_{l},uz_{j})}{G_{\nu }^{(ab)}(uz_{a},uz_{b})}\right] .
\label{zetajbets4}
\end{eqnarray}%
For $-D_{1}-1<{\mathrm{Re}}\,s<-D_{1}$ the second integral on the right is
finite in the limit $\rho \rightarrow 0$, whereas the first integral
vanishes as $\rho ^{\alpha _{s}+1}$. As a result, by using the formula $%
\Gamma (-\alpha _{s})\sin \pi \alpha _{s}=-\pi /\Gamma (\alpha _{s}+1)$, in
this strip of complex plane $s$ we have the following integral
representation
\begin{eqnarray}
\zeta _{j\beta }(s) &=&\zeta _{j\beta }^{\mathrm{(J)}}(s)-\frac{\mu
^{-s-1}B_{j}}{\Gamma \left( -\frac{s}{2}\right) \Gamma (\alpha _{s}+1)}
\notag \\
&&\times \int_{\lambda _{\beta }}^{\infty }du\,u(u^{2}-\lambda _{\beta
}^{2})^{\alpha _{s}}\Omega _{j\nu }(uz_{a},uz_{b}),  \label{zet12}
\end{eqnarray}%
where
\begin{eqnarray}
\zeta _{j\beta }^{\mathrm{(J)}}(s) &=&-\frac{n^{(j)}\mu ^{-s-1}}{\Gamma
(-s/2)\Gamma (\alpha _{s}+1)}  \notag \\
&&\times \int_{\lambda _{\beta }}^{\infty }du\,u(u^{2}-\lambda _{\beta
}^{2})^{\alpha _{s}}\frac{F_{\nu }(uz_{j})}{\bar{F}_{\nu }^{(j)}(uz_{j})}.
\label{FLs}
\end{eqnarray}%
In the last formula we use the notation $F=K$ for $j=a$ and $F=I$ for $j=b$.
The contribution of the second term on the right of Eq. (\ref{zet12}) is
finite at $s=-1$ and vanishes in the limits $z_{a}\rightarrow 0$ or $%
z_{b}\rightarrow \infty $. The first term corresponds to the contribution of
a single brane at $z=z_{j}$ when the second brane is absent. The surface
energy density corresponding to this term is located on the surface $y=a+0$
for the brane at $y=a$ and on the surface $y=b-0$ for the brane at $y=b$. To
distinguish on which side of the brane is located the corresponding term we
use the superscript J=L for the left side and J=R for the right side. As we
consider the region between the branes, in formula (\ref{FLs}) J=R for $j=a$
and J=L for $j=b$. The further analytic continuation is needed for the
function $\zeta _{j\beta }^{\mathrm{(J)}}(s)$ only and this is done in the
next section.

In the discussion above we have assumed that there is no zero mode
corresponding to $m_{n}=0$. For special values of the parameters of the
model, satisfying the condition%
\begin{equation}
(A_{a}+B_{a}\nu )(A_{b}-B_{b}\nu )\left( \frac{z_{a}}{z_{b}}\right) ^{2\nu
}=(A_{a}-B_{a}\nu )(A_{b}+B_{b}\nu ),  \label{zeromodecond}
\end{equation}%
the value $m_{n}=0$ is a solution to the eigenvalue equation (\ref{cnu}) and
the zero mode is present. We will denote this mode by $m_{0}=0$. The
corresponding contribution to the VEV of the surface energy-momentum tensor
and to the zeta function defined above is taken into account if we assume
that the summation over $n$ in the corresponding formulas includes the
summand with $n=0$ as well. The expression for this term is obtained from
the formula for general $n$ taking the limit $m_{n}\rightarrow 0$. In
particular, the contribution of the zero mode to the partial zeta function (%
\ref{zetsx}) is given by%
\begin{equation}
\zeta _{j\beta }^{(0)}(s)=\frac{\Gamma (-\alpha _{s})}{\Gamma (-s/2)\mu
^{s+1}}\frac{\lambda _{\beta }^{2}{}^{\alpha _{s}}g_{\nu }^{(l)}(0,0)}{\frac{%
\partial ^{2}}{\partial u^{2}}g_{\nu }^{(ab)}(uz_{a},uz_{b})|_{u=0}}.
\label{zeromodezet}
\end{equation}%
As before, the contribution of the nonzero modes, can be presented as the
contour integral given by Eq. (\ref{intzetsx1}), but now the contour $C$
contains an additional semicircle $C_{\rho }^{(0)}$\ of small radius which
avoids the origin in the right half-plane. Consequently, this contribution
is given by the right hand side of Eq. (\ref{zetajbets4}) plus the integral
over the contour $C_{\rho }^{(0)}$. Evaluating the latter integral we can
see that it cancels the contribution coming from the zero mode presented by
Eq. (\ref{zeromodezet}). Hence, we conclude that the integral representation
for the partial zeta function given by Eqs. (\ref{zet12}), (\ref{FLs}) is
valid for the case of the presence of the zero mode as well.

\section{Energy density on a single brane}

\label{sec:1brane}

In this section we will consider the geometry of a single brane placed at $%
y=j$. The orbifolded version of this model corresponds to the higher
dimensional generalization of the Randall-Sundrum 1-brane model with the
brane location at the orbifold fixed point $y=0$. The corresponding partial
zeta function is given by Eq. (\ref{FLs}). Now in this formula $\mathrm{J=L,R%
}$ for the left and right sides of the brane, respectively, and $F=K$ for $%
\mathrm{J=R}$ and $F=I$ for $\mathrm{J=L}$. In addition, the replacement $%
n^{(j)}\rightarrow n^{\mathrm{(J)}}$ should be done with $n^{\mathrm{(R)}}=1$
and $n^{\mathrm{(L)}}=-1$. The integral representation (\ref{FLs}) for a
single brane partial zeta function is valid in the strip $-D_{1}-1<\mathrm{Re%
}\,s<-D_{1}$ and under the assumption that the function $\bar{F}_{\nu
}^{(j)}(u)$ has no real zeros. For the analytic continuation to $s=-1$ we
employ the asymptotic expansions of the modified Bessel functions for large
values of the argument \cite{abramowiz}. For $B_{j}\neq 0$ from these
expansions one has
\begin{equation}
\frac{F_{\nu }(u)}{\bar{F}_{\nu }^{(j)}(u)}\sim \frac{1}{B_{j}}%
\sum_{l=1}^{\infty }\frac{v_{l}^{(F,j)}}{u^{l}},  \label{Inuratas}
\end{equation}%
where the coefficients $v_{l}^{(F,j)}(\nu )$ are combinations of the
corresponding coefficients in the expansions for the functions $F_{\nu }(u)$
and $F_{\nu }^{\prime }(u)$. Note that one has the relation%
\begin{equation}
v_{l}^{(K,j)}=(-1)^{l}v_{l}^{(I,j)},  \label{lelvl}
\end{equation}%
assuming that the coefficients in the boundary conditions are the same for
both sides of the brane. For the nonzero modes along the internal space $%
\Sigma $\ we subtract and add to the integrand in (\ref{FLs}) the $N$
leading terms of the corresponding asymptotic expansion and exactly
integrate the asymptotic part:
\begin{eqnarray}
\zeta _{j\beta }^{\mathrm{(J)}}(s)&=&\frac{-n^{\mathrm{(J)}}z_{j}^{-2\alpha
_{s}-2}}{\Gamma (-s/2)\mu ^{s+1}}\left[ \int_{\lambda _{\beta
}z_{j}}^{\infty }du\,u\frac{(u^{2}-\lambda _{\beta }^{2}z_{j}^{2})^{\alpha
_{s}}}{\Gamma (\alpha _{s}+1)}\right.  \notag \\
&&\times \left( \frac{F_{\nu }(u)}{\bar{F}_{\nu }^{(j)}(u)}-\sum_{l=1}^{N}%
\frac{v_{l}^{(F,j)}}{B_{j}u^{l}}\right) +\sum_{l=1}^{N}\frac{v_{l}^{(F,j)}}{%
2B_{j}}  \notag \\
&&\left. \times \frac{\left( \lambda _{\beta }z_{j}\right) ^{D_{1}+s-l+1}}{%
\Gamma (l/2)}\Gamma \left( \frac{l}{2}-\alpha _{s}-1\right) \right] .
\label{zetaLabet}
\end{eqnarray}%
For the zero mode we first separate the integral over the interval $(0,1)$
and apply the described procedure to the integral over $(1,\infty )$. By
using these formulas for $\zeta _{j\beta }^{\mathrm{(J)}}(s)$, the
corresponding function
\begin{equation}
\Phi _{j}^{\mathrm{(J)}}(s)=\frac{z_{j}^{D}B_{j}}{(4\pi )^{(D_{1}-1)/2}}%
\sum_{\beta }|\psi _{\beta }(X)|^{2}\zeta _{j\beta }^{\mathrm{(J)}}(s),
\label{FLsa}
\end{equation}%
is written in the form
\begin{widetext}
\begin{eqnarray}
\Phi _{j}^{\text{(J)}}(s) &=&-\frac{(4\pi
)^{(1-D_{1})/2}n^{\text{(J)}}z_{j}^{D_{2}}}{\Gamma \left(
-\frac{s}{2}\right) \Gamma (\alpha _{s}+1)(\mu
z_{j})^{s+1}}\Bigg\{ \sum_{\beta }|\psi _{\beta }(X)|^{2}\Bigg[
\delta _{0\lambda _{\beta }}B_{j}\int_{0}^{1}du\,u^{D_{1}+s}\frac{F_{\nu }(u)%
}{\bar{F}_{\nu }^{(j)}(u)}  \nonumber \\
&&+\int_{u_{\beta }}^{\infty
}du\,u(u^{2}-\lambda _{\beta }^{2}z_{j}^{2})^{\alpha _{s}}\left( B_{j}\frac{%
F_{\nu }(u)}{\bar{F}_{\nu }^{(j)}(u)}-\sum_{l=1}^{N}\frac{v^{F,j}_{l}}{%
u^{l}}\right) \Bigg] -
\sum_{l=1}^{N}\frac{|\psi _{0}(X)|^{2}v^{(F,j)}_{l}}{D_{1}+s-l+1}  \nonumber \\
&& +\frac{1}{2}\Gamma (\alpha
_{s}+1)\sum_{l=1}^{N}\frac{v^{(F,j)}_{l}z_{j}{}^{D_{1}+s-l+1}}{\Gamma
(l/2)}\Gamma \left( \frac{l}{2}-\alpha _{s}-1\right)\zeta _{\Sigma
}\left( \frac{l}{2}-\alpha _{s}-1,X\right) \Bigg\} , \label{FLs1}
\end{eqnarray}%
\end{widetext}where $u_{\beta }=\lambda _{\beta }z_{j}+\delta _{0\lambda
_{\beta }}$ and $\alpha _{s}$ is defined by formula (\ref{alfas}). In Eq.~(%
\ref{FLs1}) we have introduced the local spectral zeta function associated
with the massless laplacian defined on the internal subspace $\Sigma $:
\begin{equation}
\zeta _{\Sigma }(z,X)=\sideset{}{'}{\sum}_{\beta }|\psi _{\beta
}(X)|^{2}\lambda _{\beta }^{-2z},  \label{localzeta}
\end{equation}%
where the prime on the summation sign means that the zero mode should be
omitted. Both integrals in Eq. (\ref{FLs1}) are finite at $s=-1$ for $%
N\geqslant D_{1}-1$. For large values $\lambda _{\beta }$ the second
integral behaves as $\lambda _{\beta }^{D_{1}+s-N}$ and the series over $%
\beta $ in Eq. (\ref{FLs1}) is convergent at $s=-1$ for $N>D-1$. For these
values $N$ the poles at $s=-1$ are contained only in the last two terms on
the right.

The zero mode part has a simple pole at $s=-1$ presented by the summand $%
l=D-1$ of the second sum in figure braces. The pole part corresponding to
the nonzero modes is extracted from the pole structure of the local zeta
function (\ref{localzeta}). The latter is given by the formula%
\begin{equation}
\Gamma (z)\zeta _{\Sigma }(z,X)|_{z=p}=\frac{C_{D_{2}/2-p}(X)}{z-p}+\Omega
_{p}(X)+\mathcal{\cdots },  \label{zetapoles}
\end{equation}%
where $p$ is a half integer, the coefficients $C_{D_{2}/2-p}(X)$ are related
to the Seeley-DeWitt or heat kernel coefficients for the corresponding
non-minimal laplacian, and the dots denote the terms vanishing at $z=p$. In
the way similar to that used in Ref. \cite{Flac03}, it can be seen that the
coefficients $C_{p}(X)$ are related to the corresponding coefficients $%
C_{p}(X,m)$ for the massive zeta function%
\begin{equation}
\zeta _{\Sigma }(s,X;m)=\sum_{\beta }\frac{|\psi _{\beta }(X)|^{2}}{(\lambda
_{\beta }^{2}+m^{2})^{s}},  \label{zetSigsXm}
\end{equation}%
by the formula%
\begin{equation}
C_{p}(X)=C_{p}(X,0)-|\psi _{0}(X)|^{2}\delta _{p,D_{2}/2}.  \label{CpX}
\end{equation}%
The VEV of the energy density on a single brane is derived from%
\begin{equation}
\varepsilon _{j}^{\mathrm{(J)}}=-n^{\mathrm{(J)}}k_{D}^{D}C_{j}\Phi _{j}^{%
\mathrm{(J)}}(s)|_{s=-1}.  \label{epsjJ}
\end{equation}%
By using relation (\ref{zetapoles}), the energy density is written as a sum
of pole and finite parts:
\begin{equation}
\varepsilon _{j}^{\mathrm{(J)}}=\varepsilon _{j,\mathrm{p}}^{\mathrm{(J)}%
}+\varepsilon _{j,\mathrm{f}}^{\mathrm{(J)}}.  \label{phi2Lpf}
\end{equation}%
Laurent-expanding the expression on the right of Eq.~(\ref{FLs1}) near $s=-1$%
, one finds%
\begin{eqnarray}
\varepsilon _{j,\text{\textrm{p}}}^{\text{\textrm{(J)}}} &=&-\frac{%
2k_{D}^{D}C_{j}}{(4\pi )^{D_{1}/2}(s+1)}\sum_{l=1}^{D}\frac{%
v_{l}^{(F,j)}z_{j}{}^{D-l}}{\Gamma (l/2)}  \notag \\
&&\times \left[ C_{(D-l)/2}(X)+|\psi _{0}(X)|^{2}\delta _{lD_{1}}\right]
\label{phi2Lpf1}
\end{eqnarray}%
for the pole part, and
\begin{widetext}
\begin{eqnarray}
\varepsilon _{j,\text{f}}^{\text{(J)}}
&=&2k_{D}^{D}z_{j}^{D_{2}}C_{j}\beta
_{D_{1}}\sum_{\beta }|\psi _{\beta
}(X)|^{2}\left[ \delta _{0\lambda _{\beta }}B_{j}\int_{0}^{1}du\,u^{D-1}%
\frac{F_{\nu }(u)}{\bar{F}_{\nu }^{(j)}(u)}\right.   \nonumber \\
&&\left. +\int_{u_{\beta }}^{\infty
}du\,u(u^{2}-\lambda _{\beta }^{2}z_{j}^{2})^{D_{1}/2-1}\left( B_{j}\frac{%
F_{\nu }(u)}{\bar{F}_{\nu }^{(j)}(u)}-\sum_{l=1}^{N}\frac{v^{(F,j)}_{l}}{%
u^{l}}\right) \right]   \nonumber \\
&&-k_{D}^{D}z_{j}^{D_{2}}C_{j}\beta _{D_{1}}|\psi
_{0}(X)|^{2}\Bigg\{ \sideset{}{'}{\sum}
_{l=1}^{N}\frac{2v^{(F,j)}_{l}}{D_{1}-l}%
-v^{(F,j)}_{D_{1}}\left[ 2\ln (\mu z_{j})+\psi \left(
\frac{D_{1}}{2}\right)
-\psi \left( \frac{1}{2}\right) \right] \Bigg\}   \nonumber \\
&&+\frac{k_{D}^{D}C_{j}}{(4\pi )^{D_{1}/2}}\sum_{l=1}^{N}\frac{%
v^{(F,j)}_{l}}{\Gamma (l/2)}z_{j}^{D-l}\left\{
C_{(D-l)/2}(X)\left[ 2\ln \mu -\psi \left( \frac{1}{2}\right)
\right] +\Omega _{(l-D_{1})/2}(X)\right\}   \label{phi2Lpf2}
\end{eqnarray}%
\end{widetext}for the finite part, with $\psi (x)$ being the diagamma
function. In this formula the prime on the summation sign means that the
term with $l=D_{1}$ should be omitted and it is understood that $C_{p}(X)=0$
for $p<0$. In the pole part the second term in the square brackets comes
from the zero mode along $\Sigma $ and this term is cancelled by the delta
term on the right of Eq. (\ref{CpX}). For a one parameter internal manifold $%
\Sigma $ with the length scale $L$ one has $\lambda _{\beta }\sim 1/L$ and $%
\psi _{\beta }(X)\sim 1/L^{D_{2}}$. In this case instead of the zeta
function (\ref{localzeta}) we could introduce the dimensionless function $%
\tilde{\zeta}_{\Sigma }(s,X)=L^{D_{2}-2s}\zeta _{\Sigma }(s,X)$. The
corresponding dimensionless coefficients $\tilde{C}_{D_{2}/2-p}(X)$ and $%
\tilde{\Omega}_{p}(X)$ in the formula analog to Eq. (\ref{zetapoles}), are
related to the coefficients in Eq. (\ref{zetapoles}) by the formulas%
\begin{equation}
\begin{split}
& C_{D_{2}/2-p}(X)=L^{2p-D_{2}}\tilde{C}_{D_{2}/2-p}(X), \\
& \Omega _{p}(X)=L^{2p-D_{2}}\left[ \tilde{\Omega}_{p}(X)+2\tilde{C}%
_{D_{2}/2-p}(X)\ln L\right] ,
\end{split}%
\end{equation}%
and do not depend on $L$. Now we see that the term with $\ln L$ is combined
with the term $\ln \mu $ of Eq. (\ref{phi2Lpf2}) in the form $\ln (\mu L)$.

The renormalization of the surface energy density can be done modifying the
procedure used previously for the renormalization of the Casimir energy in
the Randall-Sundrum model \cite{Flac01b,Gold00,Garr01} and in its
higher-dimensional generalizations with compact internal spaces \cite%
{Flac03,Flac03b}. The form of the counterterms needed for the
renormalization is determined by the pole part of the surface energy density
given by Eq. (\ref{phi2Lpf1}). For an internal manifold with no boundaries,
this part has the structure $\sum_{l=0}^{[(D-1)/2]}a_{l}^{\mathrm{(s)}%
}(z_{j}/L)^{2l}$. By taking into account that the intrinsic scalar curvature
$R_{j}$ for the brane at $y=j$ contains the factor $(z_{j}/L)^{2}$, we see
that the pole part can be absorbed by adding to the brane action
counterterms of the form%
\begin{equation}
\int d^{D}x\sqrt{|h|}\sum_{l=0}^{[(D-1)/2]}b_{l}^{\mathrm{(s)}}R_{j}^{l},
\label{Counterterms}
\end{equation}%
where the square brackets in the upper limit of summation mean the integer
part of the enclosed expression. By taking into account that there is the
freedom to perform finite renormalizations, we see that the renormalized
surface energy density on a single brane is given by the formula%
\begin{equation}
\varepsilon _{j}^{\mathrm{(J,ren)}}=\varepsilon _{j,\mathrm{f}}^{\mathrm{(J)}%
}+\sum_{l=0}^{[(D-1)/2]}c_{l}^{\mathrm{(s)}}(z_{j}/L)^{2l}.  \label{epsjren}
\end{equation}%
The coefficients $c_{l}^{\mathrm{(s)}}$ in the finite renormalization terms
are not computable within the framework of the model under consideration and
their values should be fixed by additional renormalization conditions.

The total surface energy density for a single brane at $y=j$ is obtained by
summing the contributions from the left and right sides:%
\begin{equation}
\varepsilon _{j}^{{\mathrm{(LR)}}}=\varepsilon _{j}^{\mathrm{(L)}%
}+\varepsilon _{j}^{\mathrm{(R)}}.  \label{epsLR}
\end{equation}%
In formulas (\ref{phi2Lpf1}), (\ref{phi2Lpf2}) we should take $F=I$ for $%
\mathrm{J=L}$ and $F=K$ for $\mathrm{J=R}$. Now we see that, assuming the
same boundary conditions on both sides of the brane, the coefficients $%
C_{p}(X,0)$ enter into the sum of pole terms in the form%
\begin{equation}
2\sum_{l=1}^{[D/2]}\frac{v_{2l}^{(I,j)}}{\Gamma (l)}%
z_{j}{}^{D-2l}C_{D/2-l}(X,0).  \label{Cinsumpole}
\end{equation}%
If the internal manifold contains no boundaries and $D$ is an odd number,
one has $C_{D/2-l}(X,0)=0$ and, hence, the pole parts coming from the left
and right sides cancel out. In this case for the total surface energy
density one obtains the formula%
\begin{widetext}
\begin{eqnarray}
\varepsilon _{j}^{\text{(LR)}}
&=&2k_{D}^{D}z_{j}^{D_{2}}C_{j}^{\text{(J)}}\beta _{D_{1}}\Bigg\{
\sum_{\beta }|\psi _{\beta }(X)|^{2}B_{j}\Bigg[ \delta
_{0\lambda _{\beta }}\int_{0}^{1}du\,u^{D-1}\left( \frac{I_{\nu }(u)}{%
\bar{I}_{\nu }^{(j)}(u)}+\frac{K_{\nu }(u)}{\bar{K}_{\nu }^{(j)}(u)}\right)
 \nonumber \\
&& +\int_{u_{\beta }}^{\infty }du\,\,u(u^{2}-\lambda _{\beta
}^{2}z_{j}^{2})^{D_{1}/2-1}\Bigg( \frac{I_{\nu }(u)}{\bar{I}_{\nu
}^{(j)}(u)}+\frac{K_{\nu }(u)}{\bar{K}_{\nu
}^{(j)}(u)} -\frac{2}{B_{j}}\sum_{l=1}^{[N/2]}\frac{v^{(I,j)}_{2l}}{u^{2l}}%
\Bigg) \Bigg]   \nonumber \\
&& -2|\psi _{0}(X)|^{2}\sum_{l=1}^{[N/2]}\frac{v^{(I,j)}_{2l}}{%
D_{1}-2l}+\Gamma \left( \frac{D_{1}}{2}\right) \sum_{l=1}^{[N/2]}\frac{%
v^{(I,j)}_{2l}}{\Gamma (l)}\frac{\Omega _{l-D_{1}/2}(X)}{%
z_{j}^{2l-D_{1}}}\Bigg\} ,  \label{epsLR1}
\end{eqnarray}%
\end{widetext}where the coefficients $v_{l}^{(I,j)}$ are defined by relation
(\ref{Inuratas}). Note that this quantity does not depend on the
renormalization scale $\mu $. In the orbifolded version of the model with a
single brane at $y=0$, which corresponds to the higher dimensional
generalization of the Randall-Sundrum 1-brane model, the bulk is symmetric
under the reflection $y\rightarrow -y$. In this model the surface densities
on the left and right sides of the brane are the same and coincide with $%
\varepsilon _{j}^{\mathrm{(R)}}$. In particular, here the abovementioned
cancellation of the pole parts from left and right sides does not take place.

For a one parameter internal space of size $L$ the surface energy density on
the brane at $y=j$ is a function on the ratio $L/z_{j}$ only. Note that in
the case of the AdS bulk the corresponding quantity does not depend on the
brane position. To discuss the physics from the point of view of an observer
residing on the brane, it is convenient to introduce rescaled coordinates%
\begin{equation}
x^{\prime M}=e^{-k_{D}j}x^{M},\;M=0,1,\ldots ,D-1.  \label{scaledCoord}
\end{equation}%
With this coordinates the warp factor \ in the metric is equal to 1 on the
brane and they are physical coordinates for an observer on the brane. For
this observer the physical size of the subspace $\Sigma $ is $%
L_{j}=Le^{-k_{D}j}$ and the corresponding KK masses are rescaled by the warp
factor: $\lambda _{\beta }^{(j)}=\lambda _{\beta }e^{k_{D}j}$. Now we see
that the surface energy density is a function on the ratio $L_{j}/(1/k_{D})$
of the physical size for the internal space (for an observer residing on the
brane) to the AdS curvature radius.

As an application of the general results presented above, we can consider a
simple example with $\Sigma =S^{1}$. In this case the bulk corresponds to
the $\mathrm{AdS}_{D+1}$ spacetime with one compactified dimension $X$. The
corresponding normalized eigenfunctions and eigenvalues are as follows
\begin{equation}
\psi _{\beta }(X)=\frac{1}{\sqrt{L}}e^{2\pi i\beta X/L},\quad \beta =0,\pm
1,\pm 2,\ldots ,  \label{psiS1}
\end{equation}%
where $L$ is the length of the compactified dimension. The surface energy
density induced on the brane is obtained from general formulas by the
replacements
\begin{equation}
\sum_{\beta }|\psi _{\beta }(X)|^{2}\rightarrow \frac{2}{L}%
\sideset{}{'}{\sum}_{\beta =0}^{\infty },\quad \lambda _{\beta }\rightarrow
\frac{2\pi }{L}|\beta |,\quad D_{2}=1,  \label{replaceforS1}
\end{equation}%
where the prime means that the summand $\beta =0$ should be taken with the
weight 1/2. For the local zeta function from Eq. (\ref{localzeta}) one has%
\begin{equation}
\zeta _{\Sigma }(s,X)=\frac{2}{L}\sum_{\beta =1}^{\infty }\left( \frac{2\pi
}{L}\beta \right) ^{-2s}=\frac{2L^{2s-1}}{(2\pi )^{2s}}\zeta _{R}(2s),
\label{zetSigXS1}
\end{equation}%
where $\zeta _{R}(z)$ is the Riemann zeta function. Now the only poles of
the function $\Gamma (z)\zeta _{\Sigma }(z,X)$ are the points $z=0,1/2$. By
using the standard formulas for the gamma function and the Riemann zeta
function (see, for instance, \cite{abramowiz}), it can be seen that one has%
\begin{equation}
C_{1/2}(X)=-1/L,\;C_{0}(X)=1/2\sqrt{\pi },  \label{ResS1}
\end{equation}%
for the residues appearing in (\ref{zetapoles}) and

\begin{equation}  \label{OmegaS1}
\begin{split}
& \Omega _{0}(X)=\frac{\gamma -2\ln L}{L},\;\Omega _{\frac{1}{2}}(X)=\frac{%
\gamma +2\ln (L/4\pi )}{2\sqrt{\pi }}, \\
& \Omega _{p}(X)=\frac{2L^{2p-1}}{(2\pi )^{2p}}\Gamma (p)\zeta
_{R}(2p),\;p\neq 0,\;\frac{1}{2},
\end{split}%
\end{equation}%
for the finite parts, with $\gamma $ being the Euler constant.

\section{Two-brane geometry and induced cosmological constant}

\label{sec:2brane}

As it has been shown in Section \ref{sec:zetafunc}, the partial zeta
function related to the surface energy density on the brane at $y=j$ is
presented in the form (\ref{zet12}), where the second term on the right is
finite at the physical point $s=-1$. By taking into account formulas (\ref%
{Fjs2}), (\ref{IFs0}), for two-brane geometry the VEV of the surface energy
density on the brane at $y=j$ is presented as the sum
\begin{equation}
\varepsilon _{j}^{{\mathrm{(s)}}}=\varepsilon _{j}^{{\mathrm{(J)}}}+\Delta
\varepsilon _{j}^{{\mathrm{(s)}}}.  \label{emt2pl2}
\end{equation}%
The first term on the right is the energy density induced on a single brane
when the second brane is absent. The second term is induced by the presence
of the second brane and is given by the formula
\begin{eqnarray}
\Delta \varepsilon _{j}^{{\mathrm{(s)}}}
&=&2C_{j}n^{(j)}(k_{D}z_{j})^{D}B_{j}^{2}\beta _{D_{1}}\sum_{\beta }|\psi
_{\beta }(X)|^{2}\int_{\lambda _{\beta }}^{\infty }du\,u  \notag \\
&&\times (u^{2}-\lambda _{\beta }^{2})^{D_{1}/2-1}\Omega _{j\nu
}(uz_{a},uz_{b}).  \label{emt2pl3}
\end{eqnarray}%
By taking into account relation (\ref{tauj}) between the VEV of the field
square and the surface energy-momentum tensor, this formula can aslo be
obtained from the last term of Eq. (\ref{phi2bulk}) evaluated at the brane $%
z=z_{j}$. As we consider the region $a\leqslant y\leqslant b$, the energy
desnity (\ref{emt2pl3}) is located on the surface $y=a+0$ for the left brane
and on the surface $y=b-0$ for the right brane. Consequently, in formula (%
\ref{emt2pl2}) we take $\mathrm{J=R}$ for $j=a$ and $\mathrm{J=L}$ for $j=b$%
. The energy densities on the surfaces $y=a-0$ and $y=b+0$ are the same as
for the corresponding single brane geometry. The expression on the right of
Eq. (\ref{emt2pl3}) is finite for all nonzero distances between the branes
and is not touched by the renormalization procedure. For a given value of
the AdS energy scale $k_{D}$ and one parameter manifold $\Sigma $ with the
length scale $L$, it is a function on the ratios $z_{b}/z_{a}$ and $L/z_{a}$%
. The first ratio is related to the proper distance between the branes,
\begin{equation}
z_{b}/z_{a}=e^{k_{D}(b-a)},  \label{zbza}
\end{equation}%
and the second one is the ratio of the size of the internal space, measured
by an observer residing on the brane at $y=a$, to the AdS curvature radius $%
k_{D}^{-1}$. The expression (\ref{emt2pl3}) for the surface energy density
can be presented in another equivalent form:
\begin{eqnarray}
\Delta \varepsilon _{j}^{{\mathrm{(s)}}} &=&\sum_{\beta }|\psi _{\beta
}(X)|^{2}\int_{\lambda _{\beta }}^{\infty }du\,u(u^{2}-\lambda _{\beta
}^{2})^{D_{1}/2-1}  \notag \\
&&\times \frac{2k_{D}^{D}z_{j}^{D+1}B_{j}^{2}\beta _{D_{1}}C_{j}}{%
B_{j}^{2}(u^{2}z_{j}^{2}+\nu ^{2})-A_{j}^{2}}  \notag \\
&&\times \frac{\partial }{\partial z_{j}}\ln \left\vert 1-\frac{\bar{I}_{\nu
}^{(a)}(uz_{a})\bar{K}_{\nu }^{(b)}(uz_{b})}{\bar{I}_{\nu }^{(b)}(uz_{b})%
\bar{K}_{\nu }^{(a)}(uz_{a})}\right\vert ,  \label{Deltepsjnew}
\end{eqnarray}%
which will be used below in the discussion of the relations between the bulk
and surface energy densities.

For the comparison with the case of the bulk spacetime $\mathrm{AdS}%
_{D_{1}+1}$ when the internal space is absent, it is useful in addition to
the VEV (\ref{emt2pl3}) to consider the corresponding quantity integrated
over the subspace~$\Sigma $:
\begin{eqnarray}
\Delta \varepsilon _{D_{1}j}^{{\mathrm{(s)}}} &=&\int_{\Sigma }d^{D_{2}}X%
\sqrt{\gamma }\,\Delta \varepsilon _{j}^{{\mathrm{(s)}}}e^{-D_{2}k_{D}j}
\notag \\
&=&e^{-D_{2}k_{D}j}\sum_{\beta }\Delta \varepsilon _{j\beta }^{{\mathrm{(s)}}%
},  \label{phi2integrated}
\end{eqnarray}%
where $\Delta \varepsilon _{j\beta }^{{\mathrm{(s)}}}$ is defined by the
relation
\begin{equation}
\Delta \varepsilon _{j}^{{\mathrm{(s)}}}=\sum_{\beta }|\psi _{\beta
}(X)|^{2}\Delta \varepsilon _{j\beta }^{{\mathrm{(s)}}}.  \label{DeltDef}
\end{equation}%
Comparing this integrated VEV with the corresponding formula from Ref. \cite%
{Saha04surf}, we see that the contribution of the zero KK mode ($\lambda
_{\beta }=0$) in Eq. (\ref{phi2integrated}) differs from the VEV of the
energy density in the bulk $\mathrm{AdS}_{D_{1}+1}$ by the order of the
modified Bessel functions: for the latter case $\nu \rightarrow \nu _{1}$
with $\nu _{1}$ defined by Eq. (\ref{nu}) with the replacement $D\rightarrow
D_{1}$. Note that for $\zeta \leqslant \zeta _{D+D_{1}+1}$ one has $\nu
\geqslant \nu _{1}$. In particular, this is the case for minimally and
conformally coupled scalar fields.

Now we turn to the investigation of the part (\ref{emt2pl3}) in the surface
energy density in asymptotic regions of the parameters. For large values of
AdS radius compared with the interbrane distance, $k_{D}(b-a)\ll 1$, the
main contribution to the integral on the right of Eq. (\ref{emt2pl3}) comes
from large values of $uz_{a}\sim \lbrack k_{D}(b-a)]^{-1}$. Assuming that $%
\tilde{B}_{a}/(b-a)$ and $m(b-a)$ are fixed, we see that the order of the
Bessel modified functions is large. Replacing these functions by their
uniform asymptotic expansions for large values of the order \cite{abramowiz}%
, one obtains%
\begin{eqnarray}
\Delta \varepsilon _{j}^{{\mathrm{(s)}}} &\approx &2n^{(j)}(1-4\zeta )\tilde{%
A}_{j}\tilde{B}_{j}\beta _{D_{1}}\sum_{\beta }|\psi _{\beta
}(X)|^{2}\int_{m_{\beta }}^{\infty }du\,u^{2}  \notag \\
&&\times \frac{(u^{2}-m_{\beta }^{2})^{\frac{D_{1}}{2}-1}(\tilde{A}_{j}^{2}-%
\tilde{B}_{j}^{2}u^{2})^{-1}}{\left[ \tilde{c}_{a}(u)\tilde{c}%
_{b}(u)e^{2u(b-a)}-1\right] },  \label{EpssmallkD}
\end{eqnarray}%
where $m_{\beta }=\sqrt{m^{2}+\lambda _{\beta }^{2}}$ and we have introduced
the notation
\begin{equation}
\tilde{c}_{j}(u)=\frac{\tilde{A}_{j}-n^{(j)}\tilde{B}_{j}u}{\tilde{A}%
_{j}+n^{(j)}\tilde{B}_{j}u},\quad j=a,b.  \label{cj}
\end{equation}%
The expression on the right of Eq. (\ref{EpssmallkD}) is the corresponding
surface energy on the brane in the bulk geometry $R^{(D_{1}-1,1)}\times
\Sigma $.

For large KK masses along $\Sigma $, $z_{a}\lambda _{\beta }\gg 1$, $\lambda
_{\beta }\gg 1$, we can replace the modified Bessel functions by the
corresponding asymptotic expansions for large values of the argument. For
the contribution of a given KK mode to the leading order this gives%
\begin{eqnarray}
\Delta \varepsilon _{j\beta }^{{\mathrm{(s)}}} &\approx
&4n^{(j)}k_{D}^{D}z_{j}^{D+1}B_{j}^{2}C_{j}\beta _{D_{1}}\int_{\lambda
_{\beta }}^{\infty }du\,u^{2}  \notag \\
&&\times \frac{(A_{j}^{2}-B_{j}^{2}u^{2}z_{j}^{2})^{-1}(u^{2}-\lambda
_{\beta }^{2})^{\frac{D_{1}}{2}-1}}{%
c_{a}(uz_{a})c_{b}(uz_{b})e^{2u(z_{b}-z_{a})}-1},  \label{phi22pllargelamb}
\end{eqnarray}%
where
\begin{equation}
c_{j}(u)=\frac{A_{j}-n^{(j)}B_{j}u}{A_{j}+n^{(j)}B_{j}u},\quad j=a,b.
\label{cj1}
\end{equation}%
If in addition one has the condition $\lambda _{\beta }(z_{b}-z_{a})\gg 1$,
the dominant contribution into the $u$-integral comes from the lower limit
and we have the formula%
\begin{eqnarray}
\Delta \varepsilon _{j\beta }^{{\mathrm{(s)}}} &\approx &\frac{%
2n^{(j)}B_{j}^{2}C_{j}}{A_{j}^{2}-(\lambda _{\beta }z_{j}B_{j})^{2}}\frac{%
k_{D}^{D}z_{j}^{D+1}}{(4\pi )^{D_{1}/2}}  \notag \\
&&\times \frac{\lambda _{\beta }^{D_{1}/2+1}e^{-2\lambda _{\beta
}(z_{b}-z_{a})}}{c_{a}(\lambda _{\beta }z_{a})c_{b}(\lambda _{\beta
}z_{b})(z_{b}-z_{a})^{D_{1}/2}}.  \label{epsjlargeKK1}
\end{eqnarray}%
In particular, for sufficiently small length scale of the internal space
this formula is valid for all nonzero KK masses and the main contribution to
the surface densities comes from the zero KK mode. In the opposite limit of
large internal space, to the leading order we obtain the corresponding
result for parallel branes in $AdS_{D+1}$ bulk \cite{Saha04surf}.

For large values of the mass with $m\gg k_{D}$, $m(z_{b}-z_{a})\gg 1$, and $%
m\gg \lambda _{\beta }$, one has $\nu \sim m/k_{D}\gg 1$. Introducing a new
integration variable $u=\nu y$ and using the uniform asymptotic expansions
for the modified Bessel functions, it can be seen that the corresponding
asymptotic formula for $\Delta \varepsilon _{j\beta }^{{\mathrm{(s)}}}$ is
obtained from (\ref{epsjlargeKK1}) by the replacement $\lambda _{\beta
}\rightarrow m$. If both masses $m$ and $\lambda _{\beta }$ are of the same
order, the asymptotic behavior for $\Delta \varepsilon _{j\beta }^{{\mathrm{%
(s)}}}$ is described by Eq. (\ref{epsjlargeKK1}) replacing $\lambda _{\beta
}\rightarrow m_{\beta }$. As we could expect in this case the induced
surface densities are exponentially suppressed.

For small interbrane distances, $k_{D}(b-a)\ll 1$, which is equivalent to $%
z_{b}/z_{a}-1\ll 1$, the main contribution into the integral in Eq. (\ref%
{emt2pl3}) comes from large values $u$ and to the leading order we obtain
formula (\ref{phi22pllargelamb}). If in addition one has $\lambda _{\beta
}(z_{b}-z_{a})\ll 1$ or equivalently $\lambda _{\beta }^{(a)}(b-a)\ll 1$, we
can put in this formula $\lambda _{\beta }=0$. Assuming $(b-a)\ll |\tilde{B}%
_{j}/A_{j}|$, for $\tilde{B}_{j}\neq 0$ to the leading order one finds%
\begin{equation}
\Delta \varepsilon _{j\beta }^{{\mathrm{(s)}}}\approx -4k_{D}\sigma
_{j}n^{(j)}\frac{C_{j}\Gamma \left( \frac{D_{1}-1}{2}\right) \zeta
_{R}(D_{1}-1)}{(4\pi )^{(D_{1}+1)/2}(b-a)^{D_{1}-1}}e^{D_{2}k_{D}j},
\label{epsjsmalldist}
\end{equation}%
where $\sigma _{j}=1$ for $|B_{a}/A_{a}|,|B_{b}/A_{b}|\gg k_{D}(b-a)$, and $%
\sigma _{j}=2^{2-D}-1$ for $|B_{j}/A_{j}|\gg k_{D}(b-a)$ and $B_{l}/A_{l}=0$%
, with $l=b$ for $j=a$ and $l=a$ for $j=b$. We see that for small interbrane
distances the sign of the induced surface energy density is determined by
the coefficient $C_{j}$ and this sign is different for two cases of $\sigma
_{j}$.

Now we consider the limit $\lambda _{\beta }z_{b}\gg 1$ assuming that $%
\lambda _{\beta }z_{a}\lesssim 1$. Using the asymptotic formulas for the
Bessel modified functions containing in the argument $z_{b}$, for the
contribution of a given nonzero KK mode we find the following results%
\begin{eqnarray}
\Delta \varepsilon _{a\beta }^{{\mathrm{(s)}}} &\approx &\frac{%
k_{D}^{D}z_{a}^{D}B_{a}^{2}C_{a}}{2^{D_{1}}\pi ^{D_{1}/2-1}}\frac{(\lambda
_{\beta }/z_{b})^{D_{1}/2}e^{-2\lambda _{\beta }z_{b}}}{c_{b}(\lambda
_{\beta }z_{b})\bar{K}_{\nu }^{(a)2}(\lambda _{\beta }z_{a})},
\label{epslargezb1} \\
\Delta \varepsilon _{b\beta }^{{\mathrm{(s)}}} &\approx &-\frac{%
k_{D}^{D}z_{b}^{D+1}B_{b}^{2}C_{b}}{2^{D_{1}-1}\pi
^{D_{1}/2-1}z_{b}^{D_{1}/2}}\frac{\lambda _{\beta }^{D_{1}/2+1}e^{-2\lambda
_{\beta }z_{b}}}{(A_{b}+\lambda _{\beta }z_{b}B_{b})^{2}}  \notag \\
&&\times \frac{\bar{I}_{\nu }^{(a)}(\lambda _{\beta }z_{a})}{\bar{K}_{\nu
}^{(a)}(\lambda _{\beta }z_{a})}.  \label{epslargezb2}
\end{eqnarray}%
This limit corresponds to the interbrane distances much larger compared with
the AdS curvature radius and with the inverse KK masses measured by an
observer on the left brane: $(b-a)\gg 1/k_{D},1/\lambda _{\beta }^{(a)}$.
For a single parameter manifold $\Sigma $ with length scale $L$ and $%
(b-a)\gg L_{a}$ these conditions are satisfied for all nonzero KK modes.

In the limit $z_{a}\lambda _{\beta }\ll 1$ for fixed $z_{b}\lambda _{\beta }$%
, by using the asymptotic formulas for the modified Bessel functions for
small values of the argument and assuming $|A_{a}|\neq |B_{a}|\nu $, one
finds%
\begin{eqnarray}
\Delta \varepsilon _{a\beta }^{{\mathrm{(s)}}} &\approx &\frac{%
k_{D}^{D}z_{a}^{D+2\nu }B_{a}^{2}C_{a}\beta _{D_{1}}}{2^{2\nu -3}\Gamma
^{2}(\nu )(A_{a}-\nu B_{a})^{2}}\int_{\lambda _{\beta }}^{\infty
}du\,u^{2\nu +1}  \notag \\
&&\times (u^{2}-\lambda _{\beta }^{2})^{D_{1}/2-1}\frac{\bar{K}_{\nu
}^{(b)}(uz_{b})}{\bar{I}_{\nu }^{(b)}(uz_{b})},  \label{epssmallza1} \\
\Delta \varepsilon _{b\beta }^{{\mathrm{(s)}}} &\approx &-\frac{k_{D}^{D_{1}}%
}{c_{a}(\nu )}\left( \frac{z_{a}}{z_{b}}\right) ^{2\nu
}k_{D}^{D_{2}}z_{b}^{D_{2}}f_{\nu \beta }^{(b)},  \label{epssmallza2}
\end{eqnarray}%
where we have introduced the notation%
\begin{equation}
f_{\nu \beta }^{(b)}=\frac{4B_{b}^{2}C_{b}\beta _{D_{1}}}{2^{2\nu }\nu
\Gamma ^{2}(\nu )}\int_{\lambda _{\beta }z_{b}}^{\infty }du\,\frac{u^{2\nu
+1}}{\bar{I}_{\nu }^{(b)2}(u)}(u^{2}-\lambda _{\beta
}^{2}z_{b}^{2})^{D_{1}/2-1}.  \label{fnubet}
\end{equation}%
For $|A_{a}|=|B_{a}|\nu $ we should take into account the next terms in the
corresponding expansions of the modified Bessel functions. The integral in
Eq. (\ref{epssmallza1}) is negative for small values of the ratio $%
A_{b}/B_{b}$ and is positive for large values of this ratio. As it follows
from Eq. (\ref{epssmallza2}), for large interbtane separations the sign of
the quantity $\Delta \varepsilon _{b}^{{\mathrm{(s)}}}$ is determined by the
combination $(B_{a}^{2}\nu ^{2}-A_{a}^{2})C_{b}$ of the coefficients in the
boundary conditions. In the limit under consideration the KK masses measured
by an observer on the brane at $y=a$ are much less than the AdS energy
scale, $\lambda _{\beta }^{(a)}\ll k_{D}$, and the interbrane distance is
much larger than the AdS curvature radius. In particular, substituting $%
\lambda _{\beta }=0$, from these formulas we obtain the asymptotic behavior
for the contribution of the zero mode to the surface energy density induced
by the second brane in the limit $z_{a}/z_{b}\ll 1$. Now combining the
corresponding result with formulas (\ref{epslargezb1}), (\ref{epslargezb2}),
we see that under the conditions $(b-a)\gg 1/k_{D},L_{a}$, and $%
L_{a}k_{D}\gtrsim 1$, the contribution of the nonzero KK modes along $\Sigma
$ is suppressed with respect to the contribution of the zero mode by the
factor $(z_{b}/z_{a})^{2\nu +D_{1}/2+\delta _{j}^{b}}\exp (-2\lambda _{\beta
}z_{b})$.

From the analysis given above it follows that in the limit when the right
brane tends to the AdS horizon, $z_{b}\rightarrow \infty $, the energy
density $\Delta \varepsilon _{a\beta }^{{\mathrm{(s)}}}$ vanishes as $%
e^{-2\lambda _{\beta }z_{b}}/z_{b}^{D_{1}/2}$ for the nonzero KK mode along $%
\Sigma $ and as $z_{b}^{-D_{1}-2\nu }$ for the zero mode. The energy density
on the right brane, $\Delta \varepsilon _{b\beta }^{{\mathrm{(s)}}}$,
vanishes as $z_{b}^{D_{2}+D_{1}/2+1}e^{-2\lambda _{\beta }z_{b}}$ for the
nonzero KK mode and behaves like $z_{b}^{D_{2}-2\nu }$ for the zero mode. In
the limit when the left brane tends to the AdS boundary, $z_{a}\rightarrow 0$%
, the contribution of a given KK mode vanishes as $z_{a}^{D+2\nu }$ for $%
\Delta \varepsilon _{a\beta }^{{\mathrm{(s)}}}$ and as $z_{a}^{2\nu }$ for $%
\Delta \varepsilon _{b\beta }^{{\mathrm{(s)}}}$. For small values of the AdS
curvature radius corresponding to strong gravitational fields, assuming $%
\lambda _{\beta }z_{a}\gg 1$ and $\lambda _{\beta }(z_{b}-z_{a})\gg 1$, we
can estimate the contribution of the nonzero KK modes to the induced energy
densities by formula (\ref{epsjlargeKK1}). In particular, for the case of a
single parameter internal space with the length scale $L$, under the assumed
conditions the length scale of the internal space measured by an observer on
the brane at $y=a$ is much smaller compared to the AdS curvature radius, $%
L_{a}\ll k_{D}^{-1}$. If $L_{a}\gtrsim k_{D}^{-1}$ one has $\lambda _{\beta
}z_{a}\lesssim 1$ and to estimate the contribution of the induced surface
densities we can use formulas (\ref{epslargezb1}) and (\ref{epslargezb2}),
and the suppression is stronger compared with the previous case. For the
zero KK mode, under the condition $k_{D}(b-a)\gg 1$ we have $z_{a}/z_{b}\ll
1 $ and to the leading order the corresponding energy densities are
described by relations (\ref{epssmallza1}) and (\ref{epssmallza2}). From
these formulae it follows that the induced energy densities integrated over
the internal space behave as $k_{D}^{D_{1}+1}\exp [(D_{1}\delta
_{j}^{a}+2\nu )k_{D}(a-b)]$ for the brane at $y=j$ and are exponentially
suppressed. Note that in the model without the internal space we have
similar behavior with $\nu $ replaced by $\nu _{1}$ and for a scalar field
with $\zeta <\zeta _{D+D_{1}+1}$ the suppression is relatively weaker.

\begin{figure}[tbph]
\epsfig{figure=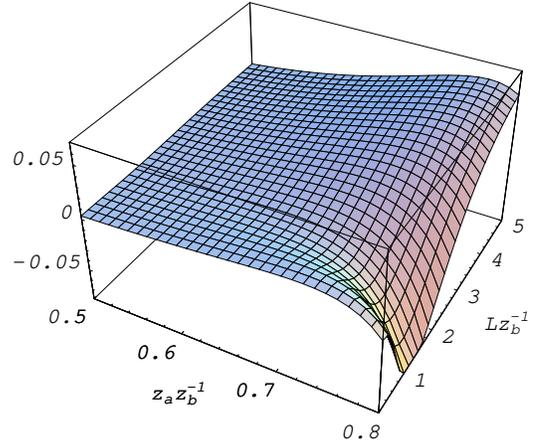,width=7cm}
\caption{Surface energy density, $\Delta \protect\varepsilon _{a}^{{\mathrm{%
(s)}}}/k_{D}^{D}$, induced on the brane at $y=a$ as a function of $%
z_{a}/z_{b}$ and $L/z_{b}$ for a $D=5$ minimally coupled massless scalar
field in the model with $\Sigma =S^{1}$ and with the Robin coefficients $%
\tilde{A}_{a}/B_{a}=-1$, $\tilde{A}_{b}/B_{b}=5$.}
\label{fig1}
\end{figure}

\begin{figure}[tbph]
\epsfig{figure=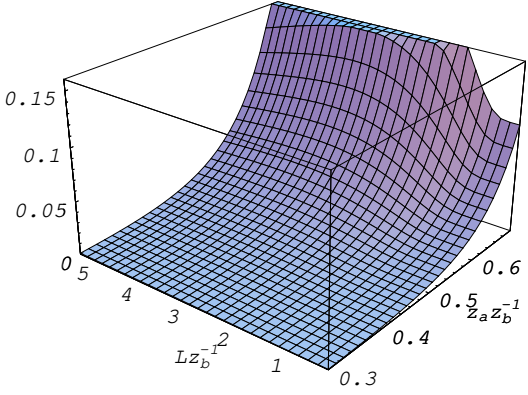,width=7cm}
\caption{The same as in Fig. \protect\ref{fig1} for the right brane.}
\label{fig2}
\end{figure}

In Fig. \ref{fig1} and Fig. \ref{fig2} we have plotted the energy densities
induced on the left and right branes, respectively, by the presence of the
second brane as functions on the size of the internal space $\Sigma =S^{1}$
and interbrane distance for a $D=5$ minimally coupled massless scalar field
with the Robin coefficients $\tilde{A}_{a}/B_{a}=-1$, $\tilde{A}_{b}/B_{b}=5$%
. The corresponding expressions are obtained from the general formulas of
this section by the subtitutions (\ref{replaceforS1}).

Introducing the rescaled coordinates defined by Eq. (\ref{scaledCoord}),
after the Kaluza-Klein reduction of the higher dimensional Hilbert action,
by the way similar to that in the Randall-Sundrum braneworld, it can be seen
that effective $D_{1}$-dimensional Newton's constant $G_{D_{1}j}$ measured
by an observer on the brane at $y=j$ and the fundamental $(D+1)$-dimensional
Newton's constant $G_{D+1}$ are related by the formula
\begin{equation}
G_{D_{1}j}=\frac{(D-2)k_{D}G_{D+1}}{V_{\Sigma j}\left[ e^{(D-2)k_{D}(b-a)}-1%
\right] }e^{(D-2)k_{D}(b-j)},  \label{GDj}
\end{equation}%
where%
\begin{equation}
V_{\Sigma j}=e^{-D_{2}k_{D}j}\int d^{D_{2}}X\sqrt{\gamma },  \label{Vsigj}
\end{equation}%
is the volume of the internal space measured by the same observer. In the
orbifolded version of the model an additional factor 2 appears in the
denominator of the expression on the right. Formula (\ref{GDj}) explicitly
shows two possibilities for the hierarchy generation by the redshift and
large volume effects. Note that the ratio of the Newton constants on the
branes, $G_{D_{1}b}/G_{D_{1}a}=e^{(2-D_{1})k_{D}(b-a)}$, is the same as in
the model without an internal space. For large interbrane distances one has $%
G_{D_{1}a}\sim k_{D}G_{D+1}/V_{\Sigma a}$, $G_{D_{1}b}\sim
k_{D}G_{D+1}e^{(2-D)k_{D}(b-a)}/V_{\Sigma b}$, and the gravitational
interactions on the brane $y=b$ are exponentially suppressed. This feature
is used in the Randall-Sundrum model to address the hierarchy problem. As we
will see below this mechanism also allows to obtain a naturally small
cosmological constant generated by the vacuum quantum fluctuations (for the
discussion of the cosmological constant problem within the framework of
braneworld models see references given in \cite{Saha04surf}).

As we have already mentioned, surface energy density (\ref{emt2pl3})
corresponds to the gravitational source of the cosmological constant type
induced on the brane at $y=j$ by the presence of the second brane. For an
observer living on the brane at $y=j$ the corresponding effective $D_{1}$%
-dimensional cosmological constant is determined by the relation
\begin{equation}
\Lambda _{D_{1}j}=8\pi G_{D_{1}j}\Delta \varepsilon _{D_{1}j}^{{\mathrm{(s)}}%
}=\frac{8\pi \Delta \varepsilon _{D_{1}j}^{{\mathrm{(s)}}}}{%
M_{D_{1}j}^{D_{1}-2}},  \label{effCC}
\end{equation}%
where $M_{D_{1}j}$ is the $D_{1}$-dimensional effective Planck mass scale
for the same observer and $\Delta \varepsilon _{D_{1}j}^{{\mathrm{(s)}}}$ is
defined by Eq. (\ref{phi2integrated}). Denoting by $M_{D+1}$ the fundamental
$(D+1)$-dimensional Planck mass, $G_{D+1}=M_{D+1}^{1-D}$, from Eq. (\ref{GDj}%
) one has the following relation%
\begin{equation}
\left( \frac{M_{D_{1}j}}{M_{D+1}}\right) ^{D_{1}-2}=\frac{%
(z_{b}/z_{a})^{D-2}-1}{(D-2)(z_{b}/z_{j})^{D-2}}\frac{V_{\Sigma j}}{k_{D}}%
M_{D+1}^{D_{2}+1},  \label{Planckhierarchy}
\end{equation}%
for the ratio of the effective and fundamental Planck scales. From the
asymptotic analysis for the induced vacuum densities given above it follows
that for large interbrane distances, $z_{b}/z_{a}\gg 1$, the contribution of
the modes with $\lambda _{\beta }z_{b}\gg 1$, $\lambda _{\beta
}z_{a}\lesssim 1$ is suppressed by the factor $(z_{b}/z_{a})^{2\nu
+D_{1}/2+\delta _{j}^{b}}\exp (-2\lambda _{\beta }z_{b})$ with respect to
the contribution of the zero mode. For the contribution of the modes with $%
z_{a}\lambda _{\beta }\ll 1$, $z_{b}\lambda _{\beta }\lesssim 1$ from Eqs. (%
\ref{epssmallza1}), (\ref{epssmallza2}) one has $\Delta \varepsilon _{j\beta
}^{{\mathrm{(s)}}}\sim k_{D}^{D}z_{a}^{D_{2}}(z_{a}/z_{b})^{2\nu +D\delta
_{j}^{a}-D_{2}}$. By using these relations, one obtains the following
estimate for the ratio of the induced cosmological constant (\ref{effCC}) to
the corresponding Planck scale quantity in the brane universe:
\begin{equation}
h_{j}\equiv \frac{\Lambda _{D_{1}j}}{8\pi G_{D_{1}j}M_{D_{1}j}^{D_{1}}}\sim
\left( \frac{z_{a}}{z_{b}}\right) ^{2\nu +D_{1}\delta _{j}^{a}}\left( \frac{%
k_{D}}{M_{D_{1}j}}\right) ^{D_{1}}.  \label{LambDj1}
\end{equation}%
By taking into account relation (\ref{Planckhierarchy}), this can also be
written in the form%
\begin{eqnarray}
h_{j} &\sim &\left( \frac{k_{D}^{D_{1}-1}}{V_{\Sigma j}M_{D+1}^{D-1}}\right)
^{D_{1}/(D_{1}-2)}  \notag \\
&&\times \exp \Big[k_{D}(a-b)\Big( 2\nu +D_{1}+\frac{D_{2}D_{1}}{D_{1}-2}%
\delta _{j}^{b}\Big) \Big].  \label{LambDjest1}
\end{eqnarray}%
For the model without an internal space this ratio is of the same order of
magnitude for both branes.

\begin{figure}[tbph]
\epsfig{figure=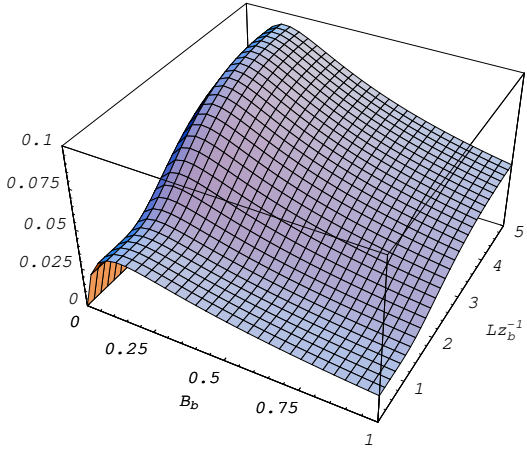,width=7cm}
\caption{Coefficient $\sum_{\protect\beta }f_{\protect\nu \protect\beta %
}^{(b)}$ in formula (\protect\ref{LambDb}) as a function on $B_b$ and $L/z_b$
for a $D=5$ minimally coupled massless scalar field in the model with $%
\Sigma =S^{1}$ and $\tilde{A}_{b}=1$.}
\label{fig3}
\end{figure}

In the higher dimensional version of the Randall-Sundrum braneworld the
brane at $z=z_{b}$ corresponds to the visible brane. For large interbrane
distances, by taking into account Eq. (\ref{epssmallza2}), for the ratio of
the induced cosmological constant (\ref{effCC}) to the Planck scale quantity
in the corresponding brane universe one obtains%
\begin{equation}
h_{b}\approx -\frac{1}{c_{a}(\nu )}\left( \frac{k_{D}}{M_{D_{1}b}}\right)
^{D_{1}}\left( \frac{z_{a}}{z_{b}}\right) ^{2\nu }\sum_{\beta }f_{\nu \beta
}^{(b)},  \label{LambDb}
\end{equation}%
where the function $f_{\nu \beta }^{(b)}$ is defined by Eq. (\ref{fnubet}).
In Fig. \ref{fig3} we have plotted the coefficient in this formula as a
function of $B_{b}$ and $L/z_{b}$ for a $D=5$ minimally coupled massless
scalar field in the model with the internal space $\Sigma =S^{1}$ and $%
\tilde{A}_{b}=1$. We recall that $L/z_{b}=k_{D}L_{b}$, where $L_{b}$ is the
physical size of the internal space for an observer living on the brane $y=b$%
.

Using relation (\ref{Planckhierarchy}) with $j=b$, we can express the
corresponding interbrane distance in terms of the ratio of the Planck scales
\begin{equation}
\frac{z_{a}}{z_{b}}\approx \left[ M_{D+1}^{D_{2}+1}\frac{V_{\Sigma b}}{k_{D}}%
\left( \frac{M_{D+1}}{M_{D_{1}b}}\right) ^{D_{1}-2}\right] ^{1/(D-2)}.
\label{Dist}
\end{equation}%
Substituting this into Eq. (\ref{LambDb}), for the ratio of the cosmological
constant on the brane at $j=b$ to the corresponding Planck scale quantity
one finds%
\begin{eqnarray}
h_{b} &\approx &-\frac{1}{c_{a}(\nu )}\left( \frac{k_{D}}{M_{D+1}}\right)
^{D_{1}-\tilde{\nu}}\left( V_{\Sigma b}M_{D+1}^{D_{2}}\right) ^{\tilde{\nu}}
\notag \\
&&\times \left( \frac{M_{D+1}}{M_{D_{1}b}}\right) ^{D_{1}+\tilde{\nu}%
(D_{1}-2)}\sum_{\beta }f_{\nu \beta }^{(b)},  \label{Lambonb}
\end{eqnarray}%
with $\tilde{\nu}=2\nu /(D-2)$. The higher dimensional Planck mass $M_{D+1}$
and AdS inverse radius $k_{D}$ are two fundamental energy scales in the
theory which in the Randall-Sundrum model are usually assumed to be of the
same order, $k_{D}\sim M_{D+1}$ (see, e.g., \cite{Ruba01}). In this case one
obtains the induced cosmological constant which is exponentially suppressed
compared with the corresponding Planck scale quantity on the visible brane.
In the model with $D_{1}=4$, $k_{D}\sim M_{D+1}\sim 1$ TeV, $M_{D_{1}b}=M_{{%
\mathrm{Pl}}}\sim 10^{16}$ TeV, assuming that the compactification scale on
the visible brane is close to the fundamental Planck scale, $V_{\Sigma
b}M_{D+1}^{D_{2}}\sim 1$, for the ratio of the induced cosmological constant
to the Planck scale quantity on the visible brane we find the estimate $%
h_{b}\sim 10^{-32(2+\tilde{\nu})}$. From (\ref{Dist}) one has $%
k_{D}(b-a)\approx 74/(D_{2}+2)$ and the corresponding interbranes distances
generating the required hierarchy between the electroweak and Planck scales
are smaller than those for the model without an internal space. In the model
proposed in Ref. \cite{Flac03b}, a separation between the fundamental Planck
scale and curvature scale is assumed: $k_{D}\sim M_{D+1}z_{a}/z_{b}\sim 1$
TeV. Under the assumption $V_{\Sigma b}M_{D+1}^{D_{2}}\sim 1$, in this model
we have $h_{b}\sim 10^{-64[1+\nu /(D+1)]}$ and $k_{D}(b-a)\approx
74/(D_{2}+3)$.

\section{Total vacuum energy and the energy balance}

\label{sec:enbal}

On background of manifolds with boundaries the total vacuum energy is
splitted into bulk and boundary parts. In the region between two branes the
bulk energy per unit coordinate volume in the $D_{1}$-dimensional subspace
is obtained by the integration of the ${_{0}^{0}}$-component of the volume
energy-momentum tensor over this region:
\begin{equation}
E^{\text{\textrm{(v)}}}=\int d^{D_{2}}Xdy\,\sqrt{|g|}\langle 0|T_{0}^{\text{%
\textrm{(v)}}0}|0\rangle .  \label{Evol}
\end{equation}%
The surface energy per unit coordinate volume in the $D_{1}$-dimensional
subspace, $E^{\text{\textrm{(s)}}}$, is related to the surface densities
evaluated in previous sections by the formula
\begin{equation}
E^{\text{\textrm{(s)}}}=\sum_{j=a,b}\frac{\varepsilon _{j}^{\text{\textrm{(s)%
}}}}{(k_{D}z_{j})^{D}}.  \label{Esurfcoord}
\end{equation}%
Now by making use of the formula for the volume energy-momentum tensor from
Ref. \cite{Saha06b} and the formula (\ref{phi2j}) for the surface densities,
it can be seen that the following formal relation takes place for the
unrenormalized VEVs:
\begin{equation}
E=E^{\text{\textrm{(v)}}}+E^{\text{\textrm{(s)}}},  \label{EvolEsurf}
\end{equation}%
where
\begin{equation}
E=\frac{1}{2}\int \frac{d^{D_{1}-1}{\mathbf{k}}}{(2\pi )^{D_{1}-1}}%
\sum_{\beta }\sum_{n=1}^{\infty }(k^{2}+m_{n}^{2}+\lambda _{\beta
}^{2})^{1/2},  \label{toten1}
\end{equation}%
is the total vacuum energy per unit coordinate volume of the $D_{1}$%
-dimensional subspace, evaluated as the sum of zero-point energies of
elementary oscillators.

The total vacuum energy within the framework of the Randall-Sundrum
braneworld is evaluated in Refs. \cite{Gold00,Toms00,Flac01b} by the
dimensional regularization method and in Ref. \cite{Garr01} by the zeta
function technique. Refs. \cite{Gold00,Toms00,Garr01} consider the case of a
minimally coupled scalar field in $D=4$, and the case of arbitrary $\zeta $
and $D$ with zero mass terms $c_{a}$ and $c_{b}$ is calculated in Ref. \cite%
{Flac01b}. For the orbifolded version of the model under consideration with $%
D_{1}=4$ and zero mass terms on the branes, the vacuum energy is
investigated in \cite{Flac03b} by using the dimensional regularization. Here
we briefly outline the zeta function approach in the general case.

We consider the zeta function related to the vacuum energy (\ref{toten1}):
\begin{equation}
\zeta (s)=\mu ^{s+1}\int \frac{d^{D_{1}-1}{\mathbf{k}}}{(2\pi )^{D_{1}-1}}%
\sum_{\beta }\sum_{n=1}^{\infty }(k^{2}+m_{n}^{2}+\lambda _{\beta
}^{2})^{-s/2},  \label{zetatot1}
\end{equation}%
where, as before, the parameter $\mu $ with dimension of mass is introduced
by dimensional reasons. The vacuum energy in the region between the branes
is obtained by the analytic continuation of this zeta function to the value $%
s=-1$:%
\begin{equation}
E=\frac{1}{2}\zeta (s)|_{s=-1}.  \label{Ezeta}
\end{equation}%
After the evaluation of the integral over ${\mathbf{k}}$ one obtains the
formula
\begin{equation}
\zeta (s)=\frac{\mu ^{s+1}}{(4\pi )^{(D_{1}-1)/2}}\frac{\Gamma \left(
s_{1}/2\right) }{\Gamma \left( s/2\right) }\sum_{\beta }\zeta _{\beta
}(s_{1}),  \label{zetatot2}
\end{equation}%
with the partial zeta function%
\begin{equation}
\zeta _{\beta }(s_{1})=\sum_{n=1}^{\infty }\left( m_{n}^{2}+\lambda _{\beta
}^{2}\right) ^{-s_{1}/2},\;s_{1}=s+1-D_{1}.  \label{zetaparten}
\end{equation}%
To evaluate the vacuum energy we need to perform the analytic continuation
of the function (\ref{zetatot2}) to the neighborhood of $s=-1$. This
corresponds to the analytic continuation for $\zeta _{\beta }(s_{1})$ to the
point $s_{1}=-D_{1}$. For this we present the partial zeta function (\ref%
{zetaparten}) as the contour integral
\begin{equation}
\zeta _{\beta }(s)=\frac{1}{2\pi i}\int_{C}du\,(u^{2}+\lambda _{\beta
}^{2})^{-s/2}\frac{d}{du}\ln g_{\nu }^{(ab)}(uz_{a},uz_{b}),
\label{zetapart2}
\end{equation}%
where the integration contour is the same as in formula (\ref{intzetsx1}).
Now by making use of the standard properties of the Bessel functions, we see
that the parts of the integrals over $(0,\pm i\lambda _{\beta })$ cancel and
we find the following integral representation%
\begin{eqnarray}
\zeta _{\beta }(s) &=&\frac{1}{\pi }\sin \frac{\pi s}{2}\int_{\lambda
_{\beta }}^{\infty }du\,(u^{2}-\lambda _{\beta }^{2})^{-s/2}  \notag \\
&&\times \frac{d}{du}\ln G_{\nu }^{(ab)}(uz_{a},uz_{b}).  \label{zetapartG}
\end{eqnarray}%
For the further discussion it is convenient to write the partial zeta
function in the decomposed form%
\begin{equation}
\zeta _{\beta }\left( s\right) =\sum_{j=a,b}\zeta _{\beta }^{(j)}\left(
s\right) +\Delta \zeta _{\beta }\left( s\right) ,  \label{zetanint2}
\end{equation}%
with the separate terms%
\begin{eqnarray}
\zeta _{\beta }^{(j)}\left( s\right) &=&\frac{1}{\pi }\sin \frac{\pi s}{2}%
\int_{\lambda _{\beta }}^{\infty }du\,(u^{2}-\lambda _{\beta }^{2})^{-s/2}
\notag \\
&&\times \frac{d}{du}\ln \left[ u^{n^{(j)}\nu }\bar{F}_{\nu }^{(j)}(uz_{j})%
\right] ,  \label{zetapartsepar} \\
\Delta \zeta _{\beta }\left( s\right) &=&\frac{1}{\pi }\sin \frac{\pi s}{2}%
\int_{\lambda _{\beta }}^{\infty }du\,(u^{2}-\lambda _{\beta }^{2})^{-s/2}
\notag \\
&&\times \frac{d}{du}\ln \left[ 1-\frac{\bar{I}_{\nu }^{(a)}(uz_{a})\bar{K}%
_{\nu }^{(b)}(uz_{b})}{\bar{K}_{\nu }^{(a)}(uz_{a})\bar{I}_{\nu
}^{(b)}(uz_{b})}\right] ,  \label{zetapartsepar1}
\end{eqnarray}%
where $F=K$ for $j=a$ and $F=I$ for $j=b$. The part $\Delta \zeta _{\beta
}\left( s\right) $ is finite at $s=-D_{1}$ and the analytic continuation is
necessary for the parts $\zeta _{\beta }^{(j)}\left( s\right) $, $j=a,b$.
The latter are the partial zeta functions for the geometry of a single brane
in the regions $y\geqslant a$ and $y\leqslant b$, respectively. On the base
of Eq. (\ref{zetanint2}) similar decomposition can be given for the total
zeta function%
\begin{equation}
\zeta \left( s\right) =\sum_{j=a,b}\zeta ^{(j)}\left( s\right) +\Delta \zeta
\left( s\right) ,  \label{zetasplit}
\end{equation}%
where separate parts are related to the corresponding partial zeta functions
by the relations similar to Eq. (\ref{zetatot2}) with $s_{1}$ defined by Eq.
(\ref{zetaparten}). In deriving the integral representation (\ref{zetapartG}%
) we have assumed that there is no zero mode corresponding to $m_{n}=0$. The
contribution of the possible zero mode to the partial zeta function $\zeta
_{\beta }(s)$ is $\lambda _{\beta }^{-s}$. As regards to the nonzero modes,
their contribution, as before, is given by the contour integral on the right
of Eq. (\ref{zetapart2}), but now the point $u=0$ has to be avoided by the
small semicircle in the right half-plane. The integrals over the parts of
the imaginary axis are transformed to the expression given by the right hand
side \ of Eq. (\ref{zetapartG}), whereas the integral over the small
semicircle is equal to $-\lambda _{\beta }^{-s}$ and cancels the
contribution from the zero mode. Hence, the integral representation (\ref%
{zetapartG}) for the partial zeta function is valid in the case of presence
of the zero mode as well.

In order to obtain an analytic continuation for the function $\zeta _{\beta
}^{(j)}(s)$, in the integrand for the non-zero modes we subtract and add the
term which is obtained replacing the function $\bar{F}_{\nu }^{(j)}(uz_{j})$
by the first $N$ terms of the corresponding asymptotic expansion for large
values $u$. In the zero mode part we split the integral in Eq. (\ref%
{zetapartsepar}) into the integrals over $(0,1)$ and $(1,\infty )$. The
contribution of the first integral to the zeta function is finite at $%
s=-D_{1}$ and the part with the second integral can be treated in the way
similar to that used for the non-zero modes. After the explicit integration
of the asymptotic part and introducing the notation
\begin{equation}
\Sigma _{\nu }^{(F,j)}(u)=-n^{(j)}\sqrt{\frac{2q_{F}}{u}}\frac{e^{n^{(j)}u}}{%
B_{j}}\bar{F}_{\nu }^{(j)}(u),  \label{SigIb}
\end{equation}%
with $q_{I}=\pi $ and $q_{K}=1/\pi $,\ for the corresponding zeta function
the following formula is obtained%
\begin{widetext}
\begin{eqnarray}
\zeta ^{(j)}(s) &=&\frac{\mu ^{s+1}(4\pi )^{(1-D_{1})/2}z_{j}^{s_{1}}}{%
\Gamma \left( s/2\right) \Gamma (1-s_{1}/2)}\Bigg\{ \sum_{\beta
}\Bigg[ \delta _{0\lambda _{\beta
}}\int_{0}^{1}du\,u^{-s_{1}}\frac{d}{du}\ln \left( \Sigma _{\nu
}^{(F,j)}(u)\right) +\int_{u_{\beta }}^{\infty }du\,(u^{2}-\lambda
_{\beta
}^{2}z_{j}^{2})^{-s_{1}/2}   \nonumber \\
&& \times \frac{d}{du}\Bigg( \ln
\left( \Sigma _{\nu }^{(F,j)}(u)\right) -\sum_{l=1}^{N}\frac{w_{l}^{(F,j)}}{%
u^{l}}\Bigg) \Bigg]  -\Gamma \left(1-\frac{s_{1}}{2}\right)
\sum_{l=-1}^{N}\frac{\tilde{w}_{l}^{(F,j)}}{z_{j}^{l+s_{1}}}%
\zeta _{\Sigma }\left( \frac{l+s_{1}}{2}\right) \Gamma \left( \frac{l+s_{1}}{%
2}\right)  \nonumber \\
&& -\sum_{l=1}^{N}l\frac{w_{l}^{(F,j)}}{s_{1}+l} \Bigg\}
\label{zetab}
\end{eqnarray}%
\end{widetext}where, as before, $u_{\beta }=\lambda _{\beta }z_{j}+\delta
_{0\lambda _{\beta }}$, and we have used the relation $\sin (\pi z)\Gamma
(1-z)=\pi /\Gamma (z)$ for the gamma function. In formula (\ref{zetab}),%
\begin{eqnarray}
&& \tilde{w}_{-1}^{(F,j)}=\frac{n^{(j)}}{2\sqrt{\pi }},\;\tilde{w}%
_{0}^{(F,j)}=-\frac{1+2n^{(j)}\nu }{4},  \notag \\
&& \tilde{w}_{l}^{(F,j)}=\frac{w_{l}^{(F,j)}}{\Gamma (l/2)},\;l=1,2,\ldots ,
\label{wtildeFj}
\end{eqnarray}
with $w_{l}^{(F,j)}$ being the coefficients in the asymptotic expansion of
the function $\ln [\Sigma _{\nu }^{(F,j)}(u)]$ for large values of the
argument:
\begin{equation}
\ln [\Sigma _{\nu }^{(F,j)}(u)]\sim \sum_{l=1}^{\infty }\frac{w_{l}^{(F,j)}}{%
u^{l}},\quad j=a,b,  \label{Defwnul}
\end{equation}%
and we have defined the global zeta function
\begin{equation}
\zeta _{\Sigma }(s)=\sideset{}{'}{\sum}_{\beta }\lambda _{\beta }^{-2s}.
\label{zetaSigglob}
\end{equation}%
Note that one has the relation $w_{l}^{(K,j)}=(-1)^{l}w_{l}^{(I,j)}$,
assuming that the coefficients in the boundary conditions are the same, and
these coefficients are related to the coefficients in the similar expansions
for the modified Bessel functions (see, for instance, \cite{abramowiz}). For
$N>D_{1}$ both integrals on the right of formula (\ref{zetab}) are finite at
the physical point $s=-1$. For large values of $\lambda _{\beta }$ the
second integral behaves as $(\lambda _{\beta }z_{j})^{-N-s_{1}}$ and, hence,
the series over $\beta $ is convergent at the point $s_{1}=-D_{1}$ if $N>D$.
As a result, for these values of $N$ the only poles in the expression (\ref%
{zetab}) are those for the function $\Gamma (z)\zeta _{\Sigma }(z)$ and the
pole in the last sum corresponding to the term with $l=D_{1}$. Note that
taking $N=\infty $ we obtain the expansion of the zeta function over $%
1/z_{j} $.

In order to separate the pole and finite parts of the zeta function $\zeta
^{(j)}(s)$ at the physical point $s=-1$, we employ the global analog of
formula (\ref{zetapoles}):
\begin{equation}
\Gamma (z)\zeta _{\Sigma }(z)|_{z=p}=\frac{C_{D_{2}/2-p}}{z-p}+\Omega _{p}+%
\mathcal{\cdots }.  \label{Gamzetglob}
\end{equation}%
As it has been shown in Ref. \cite{Flac03}, the coefficients $C_{p}$ in this
formula are related to the integrated Seeley-DeWitt coefficients $C_{p}(m)$
for the massive zeta function $\zeta _{\Sigma }(s;m)=\sum_{\beta }(\lambda
_{\beta }^{2}+m^{2})^{-s}$ by the formula%
\begin{equation}
C_{p}=C_{p}(0)-\delta _{p,D_{2}/2},  \label{CpCp0}
\end{equation}%
which is obtained from (\ref{CpX}) by the integration over the internal
space. For the model with the internal space $\Sigma =S^{1}$ the zeta
function $\zeta _{\Sigma }(z)$ is related to the corresponding local zeta
function discussed in Section \ref{sec:1brane} by the formula $\zeta
_{\Sigma }(z)=L\zeta _{\Sigma }(z,X)$ and the expressions for the
coefficients $C_{D_{2}/2-p}$ and $\Omega _{p}$ are directly obtained from (%
\ref{ResS1}), (\ref{OmegaS1}).

By making use of formula (\ref{Gamzetglob}), near the point $s=-1$ the zeta
function $\zeta ^{(j)}(s)$ is presented in the form of the sum of pole and
finite parts:%
\begin{equation}
\zeta ^{(j)}(s)|_{s=-1}=\frac{\zeta _{-1}^{(j)}}{s+1}+\zeta _{0}^{(j)},
\label{polefin}
\end{equation}%
with the residue%
\begin{equation}
\zeta _{-1}^{(j)}=\frac{2}{(4\pi )^{D_{1}/2}}\sum_{l=-1}^{N}\frac{\tilde{w}%
_{l}^{(F,j)}}{z_{j}^{l}}\left[ C_{(D-l)/2}+\delta _{lD_{1}}\right] .
\label{resE}
\end{equation}%
In formula (\ref{resE}) the part with the first term in the square brackets
comes from the nonzero KK modes along $\Sigma $ and the part with the second
term comes from the zero mode. Now by taking into account formula (\ref%
{CpCp0}), we see that the zero mode part is cancelled by the delta term \ in
Eq. (\ref{CpCp0}). For the finite part of the zeta function we obtain the
formula%
\begin{widetext}
\begin{eqnarray}
\zeta ^{(j)}_{0} &=&- \frac{2\beta
_{D_{1}}}{D_{1}z_{j}^{D_{1}}}\sum_{\beta }\Bigg\{ \delta
_{0\lambda _{\beta }}\int_{0}^{1}du\,u^{D_{1}}\frac{d}{du}\ln
\left( \Sigma
_{\nu }^{(F,j)}(u)\right)  +\int_{u_{\beta }}^{\infty }du\,(u^{2}-\lambda _{\beta
}^{2}z_{j}^{2})^{D_{1}/2}\frac{d}{du}\Bigg[ \ln
\left( \Sigma _{\nu }^{(F,j)}(u)\right) -\sum_{l=1}^{N}\frac{w_{l}^{(F,j)}}{%
u^{l}}\Bigg] \Bigg\}\nonumber \\
&& +\frac{2\beta _{D_{1}}}{D_{1}z_{j}^{D_{1}}}%
\Bigg\{
\sideset{}{'}{\sum}_{l=1}^{N}\frac{lw_{l}^{(F,j)}}{l-D_{1}}+
D_{1}w_{D_{1}}^{(F,j)}\left[ \frac{1}{2}\psi
\left(\frac{D_{1}}{2}+1\right)-\frac{1}{2}\psi \left(
-\frac{1}{2}\right) +\ln (\mu z_{j})\right] \Bigg\}
 \nonumber \\
&&+\frac{1}{(4\pi )^{D_{1}/2}} \sum_{l=-1}^{N}\frac{\tilde
w_{l}^{(F,j)}}{z_{j}^{l}}\left[ 2C_{(D-l)/2}\left( \ln \mu
-\frac{1}{2}\psi \left( -\frac{1}{2}\right) \right) +\Omega _{(
l-D_{1}) /2}\right] . \label{zetajnear-1}
\end{eqnarray}
\end{widetext}where the prime on the summation sign means that the term $%
l=D_{1}$ should be omitted. Note that taking in this formula $N=\infty $ we
obtain the expansion of the finite part over $1/z_{j}$. For an internal
space with the length scale $L$ we can introduce the dimensionless zeta
function $\tilde{\zeta}_{\Sigma }(s)=L^{-2s}\zeta _{\Sigma }(s)$ with the
coefficients $\tilde{C}_{D_{2}/2-p} $ and $\tilde{\Omega}_{p}$ defined by
the relation similar to Eq. (\ref{Gamzetglob}). Now it can be seen that in
terms of these coefficients the last term on the right of formula (\ref%
{zetajnear-1}) is rewritten as%
\begin{eqnarray}
&&\frac{L^{-D_{1}}}{(4\pi )^{D_{1}/2}}\sum_{l=-1}^{N}\tilde{w}%
_{l}^{(F,j)}\left( \frac{L}{z_{j}}\right) ^{l}\bigg[2\tilde{C}_{(D-l)/2}
\notag \\
&&\times \left( \ln (\mu L)-\frac{1}{2}\psi \left( -\frac{1}{2}\right)
\right) +\tilde{\Omega}_{(l-D_{1})/2}\bigg].  \label{Elast}
\end{eqnarray}%
This explicitly shows that the combination $\zeta _{0}^{(j)}z_{j}^{D_{1}}$
is a function on the ratio $L/z_{j}$ only$.$

Similar to the case of the surface energy, for a single brane at $y=j$ we
denote by $E_{j}^{\mathrm{(J)}}$ the total vacuum energy in the region $%
y\geqslant j$ for $\mathrm{J=R}$ and in the region $y\leqslant j$ for $%
\mathrm{J=L}$. This energy is related to the zeta function considered above
by the formula%
\begin{equation}
E_{j}^{\mathrm{(J)}}=\frac{1}{2}\zeta ^{(j)}(s)|_{s=-1},  \label{EjJ}
\end{equation}%
where in the formulas for the function $\zeta ^{(j)}(s)$ it should be taken $%
F=I$ for $\mathrm{J=L}$ and $F=K$ for $\mathrm{J=R}$. In accordance with Eq.
(\ref{polefin}) the regularized vacuum energy contains pole and finite
parts. The structure of the residue $\zeta _{-1}^{(j)}$ determines the form
of the counterterms which should be added to the action for the
renormalization of the vacuum energy. From formula (\ref{resE}) it follows
that for an internal space without boundaries the pole terms can be absorbed
by adding to the brane action the counterterm
\begin{equation}
\int d^{D}x\sqrt{|h|}\sum_{l=0}^{[(D+1)/2]}a_{l}R_{j}^{l},  \label{CountEj}
\end{equation}%
where $R_{j}$ is the intrinsic curvature scalar for the brane at $y=j$. Note
that this counterterm has the same structure as that for the surface energy.
The only difference is the additional summand corresponding to the term $%
l=[(D+1)/2]$. Hence, by adding the counterterm (\ref{CountEj}) we can absorb
the pole terms for both total and surface energies. Now by taking into
account the possibility for the appearance of the finite renormalization
terms, the renormalized vacuum energy for the geometry of a single brane is
presented in the form%
\begin{equation}
E_{j}^{\mathrm{(J,ren)}}=E_{j,\mathrm{f}}^{\mathrm{(J)}}+%
\sum_{l=0}^{[(D+1)/2]}c_{l}(z_{j}/L)^{2l},  \label{EjJren}
\end{equation}%
where $E_{j,\mathrm{f}}^{\mathrm{(J)}}=\zeta _{0}^{(j)}/2$ and the
coefficients $c_{l}$ in the finite renormalization terms should be fixed by
imposing renormalization conditions.

As in the case of the surface energies, now we see that for the internal
manifold with no boundaries in the calculation of the total vacuum energy
for a single brane at $y=j$, including the contributions from the left and
right regions, the pole parts of the energies cancel out for odd values $D$
(assuming that the coefficients in the boundary conditions (\ref{boundcond})
on the right and left surfaces are the same) and we obtain a finite result.
For this energy one receives%
\begin{widetext}
\begin{eqnarray}
E_{j} &=&-\frac{2\beta _{D_{1}}}{D_{1}z_{j}^{D_{1}}}\sum_{\beta
}\Bigg[ \delta _{0\lambda _{\beta
}}\int_{0}^{1}du\,u^{D_{1}}\frac{d}{du}\ln \left(
\Sigma _{\nu }^{(I,j)}(u)\Sigma _{\nu }^{(K,j)}(u)\right)  \notag \\
&& +\int_{u_{\beta }}^{\infty }du\,(u^{2}-\lambda _{\beta
}^{2}z_{j}^{2})^{D_{1}/2}\frac{d}{du}\Bigg( \ln \left( \Sigma
_{\nu }^{(I,j)}(u)\Sigma _{\nu }^{(K,j)}(u)\right)
-2\sum_{l=1}^{M}\frac{w_{2l}^{(I,j)}}{u^{2l}}\Bigg) \Bigg]  \notag \\
&&+\frac{4\beta _{D_{1}}}{z_{j}^{D_{1}}}\sum_{l=1}^{M}\frac{%
lw_{2l}^{(I,j)}}{D_{1}(2l-D_{1})}+\frac{1}{(4\pi
)^{D_{1}/2}}\Bigg[
\sum_{l=1}^{M}\frac{4w_{2l}^{(I,j)}}{z_{j}^{2l}\Gamma (l)}\Omega
_{l-D_{1}/2}-\Omega _{-D_{1}/2}\Bigg] . \label{Ejfinite}
\end{eqnarray}%
\end{widetext}with $M>(D-1)/2$. Note that the energy per unit physical
volume in the $D_{1}$-dimensional subspace is determined as $%
(k_{D}z_{j})^{D_{1}}E^{(j)}$ and is a function on the ratio $L/z_{j}$ only.

On the base of the analysis given above, for the geometry of two branes the
total energy in the region $a\leqslant y\leqslant b$ is presented in the
form
\begin{equation}
E=E_{a}^{\mathrm{(R)}}+E_{b}^{\mathrm{(L)}}+\Delta E,  \label{totendec}
\end{equation}%
where the interference term $\Delta E$ in this formula is finite for all
nonzero values of the interbrane separation and is obtained directly from
the part $\Delta \zeta \left( s\right) $ substituting $s=-1$. After the
integration by parts this term is presented in the form%
\begin{eqnarray}
\Delta E &=&\beta _{D_{1}}\sum_{\beta }\int_{\lambda _{\beta }}^{\infty
}du\,u(u^{2}-\lambda _{\beta }^{2})^{D_{1}/2-1}  \notag \\
&&\times \ln \left\vert 1-\frac{\bar{I}_{\nu }^{(a)}(uz_{a})\bar{K}_{\nu
}^{(b)}(uz_{b})}{\bar{K}_{\nu }^{(a)}(uz_{a})\bar{I}_{\nu }^{(b)}(uz_{b})}%
\right\vert ,  \label{DeltaEhigh}
\end{eqnarray}%
and is not affected by finite renormalizations.

Now we briefly discuss the behavior of the interference part in the vacuum
energy in the asymptotic regions of the parameters. For large values of the
AdS radius compared with the interbrane distance, $k_{D}(b-a)\ll 1$, by
making use the uniform asymptotic expansions for the modified Bessel
functions, to the leading order one finds the result for two parallel branes
on the bulk $R^{(D_{1}-1,1)}\times \Sigma $:
\begin{eqnarray}
\Delta E &\approx &\beta _{D_{1}}\sum_{\beta }\int_{m_{\beta }}^{\infty
}du\,u(u^{2}-m_{\beta }^{2})^{D_{1}/2-1}  \notag \\
&&\times \ln \left\vert 1-\frac{e^{-2u(b-a)}}{\tilde{c}_{a}(u)\tilde{c}%
_{b}(u)}\right\vert ,  \label{DeltaEsmallkD}
\end{eqnarray}%
where $m_{\beta }=\sqrt{m^{2}+\lambda _{\beta }^{2}}$ and $\tilde{c}_{j}(u)$
is defined by formula (\ref{cj}). For large KK masses along the internal
space, $z_{a}\lambda _{\beta }\gg 1$, for the contribution of the mode with
a given $\beta $ one has%
\begin{eqnarray}
\Delta E_{\beta } &\approx &\beta _{D_{1}}\int_{\lambda _{\beta }}^{\infty
}du\,u(u^{2}-\lambda _{\beta }^{2})^{D_{1}/2-1}  \notag \\
&&\times \ln \left\vert 1-\frac{e^{-2u(z_{b}-z_{a})}}{%
c_{a}(uz_{a})c_{b}(uz_{b})}\right\vert ,  \label{DeltElargeKK}
\end{eqnarray}%
with $c_{j}(u)$ defined by Eq. (\ref{cj1}). This contribution is
exponentially suppressed if $\lambda _{\beta }(z_{b}-z_{a})\gg 1$:%
\begin{equation}
\Delta E_{\beta }\approx \frac{1}{2(4\pi )^{D_{1}/2}}\left( \frac{\lambda
_{\beta }}{z_{b}-z_{a}}\right) ^{D_{1}/2}\frac{e^{-2\lambda _{\beta
}(z_{b}-z_{a})}}{c_{a}(\lambda _{\beta }z_{a})c_{b}(\lambda _{\beta }z_{b})}.
\label{DeltElargeKK1}
\end{equation}%
For small interbrane distances, satisfying the conditions $k_{D}(b-a)\ll 1$
and $\lambda _{\beta }(z_{b}-z_{a})\ll 1$ to the leading order we have the
formula%
\begin{equation}
\Delta E_{\beta }\approx -\sigma \frac{\zeta _{\mathrm{R}%
}(D_{1}+1)e^{-D_{1}k_{D}a}}{(4\pi )^{(D_{1}+1)/2}(b-a)^{D_{1}}}\Gamma \left(
\frac{D_{1}+1}{2}\right) ,  \label{DeltEsmalldist}
\end{equation}%
where $\sigma =1$ for $c_{a}(\lambda _{\beta }z_{a})c_{b}(\lambda _{\beta
}z_{b})>0$ and $\sigma =2^{-D_{1}}-1$ for $c_{a}(\lambda _{\beta
}z_{a})c_{b}(\lambda _{\beta }z_{b})<0$. Noting that the renormalized vacuum
energies for single branes are finite in the limit $a\rightarrow b$, we see
that for sufficiently small interbrane distances the totall vacuum energy is
dominated by the interference part. In particular, it follows from here that
for the case $\sigma =1$ any fixation of the interbrane distance can be
stable only locally.

In the limit $\lambda _{\beta }z_{b}\gg 1$ assuming that $\lambda _{\beta
}z_{a}\lesssim 1$, for the contribution of a given KK mode one has the
estimate%
\begin{equation}
\Delta E_{\beta }\approx -\pi \left( \frac{\lambda _{\beta }}{4\pi z_{b}}%
\right) ^{D_{1}/2}\frac{e^{-2\lambda _{\beta }z_{b}}}{c_{b}(\lambda _{\beta
}z_{b})}\frac{\bar{I}_{\nu }^{(a)}(\lambda _{\beta }z_{a})}{\bar{K}_{\nu
}^{(a)}(\lambda _{\beta }z_{a})},  \label{DeltElim1}
\end{equation}%
and the contribution from this part of KK spectrum is exponentially small.
For $\lambda _{\beta }z_{a}\ll 1$ and for fixed $\lambda _{\beta }z_{b}$ we
find%
\begin{eqnarray}
\Delta E_{\beta } &\approx &-\frac{2^{1-2\nu }\beta _{D_{1}}z_{a}^{2\nu }}{%
v\Gamma ^{2}(\nu )c_{a}(\nu )}\int_{\lambda _{\beta }}^{\infty }du\,u^{2\nu
+1}  \notag \\
&&\times (u^{2}-\lambda _{\beta }^{2})^{D_{1}/2-1}\frac{\bar{K}_{\nu
}^{(b)}(uz_{b})}{\bar{I}_{\nu }^{(b)}(uz_{b})}.  \label{DeltElim2}
\end{eqnarray}%
From the asymptotic analysis given above it follows that when the right
brane tends to the AdS horizon, $z_{b}\rightarrow \infty $, one has the
estimate $\Delta E_{\beta }\sim e^{-2\lambda _{\beta }z_{b}}/z_{b}^{D_{1}/2}$
for the nonzero KK modes and $\Delta E_{\beta }\sim z_{b}^{-D_{1}-2\nu }$
for the zero KK mode. When the left brane tends to the AdS boundary, $%
z_{a}\rightarrow 0$, the interference part of the vacuum energy behaves as $%
\Delta E\sim z_{a}^{2\nu }$. For small values of the AdS curvature radius
corresponding to strong gravitational fields the main contribution comes
from the zero mode and the interference part is suppressed by the factor $%
e^{2\nu k_{D}(a-b)}/z_{b}^{D_{1}}$.

\begin{figure}[tbph]
\epsfig{figure=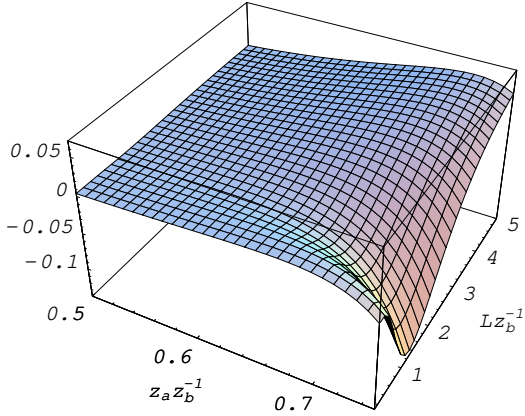,width=7cm}
\caption{Interference part of the vacuum energy, $\Delta E/k_{D}^{D+1}$, as
a function of $z_{a}/z_{b}$ and $L/z_{b}$ for a $D=5$ minimally coupled
massless scalar field in the model with $\Sigma =S^{1}$ and with the Robin
coefficients $\tilde{A}_{a}/B_{a}=-1$, $\tilde{A}_{b}/B_{b}=5$.}
\label{fig4}
\end{figure}

\begin{figure}[tbph]
\epsfig{figure=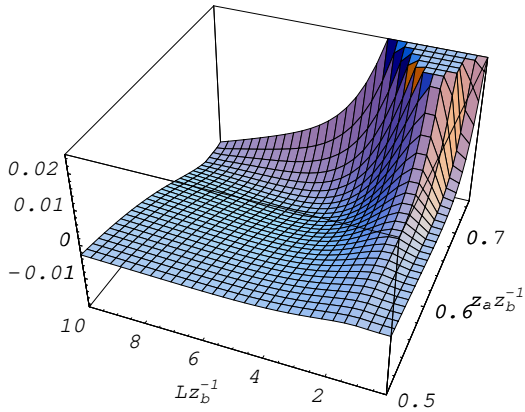,width=7cm}
\caption{The same as in Fig. \protect\ref{fig4} for the values of the Robin
coefficients $B_{a}=0$, $\tilde{A}_{b}/B_{b}=5$.}
\label{fig5}
\end{figure}

In figures we have presented the dependence of the interference part of the
vacuum energy on the size of the internal space $\Sigma =S^{1}$ and
interbrane distance for a $D=5$ minimally coupled massless scalar field with
the Robin coefficients $\tilde{A}_{a}/B_{a}=-1$, $\tilde{A}_{b}/B_{b}=5$
(Fig. \ref{fig4}) and $B_{a}=0$, $\tilde{A}_{b}/B_{b}=5$ (Fig. \ref{fig5}).
Note that, in accordance with the asymptotic analysis, for small interbrane
distances this part is negative for the first case and positive for the
second one. In the second example the interference part of the vacuum energy
has minimum with respect to both variables

Now let us check that for the separate parts of the vacuum energy the
standard energy balance equation takes places. We denote by $P$ the
perpendicular vacuum stress on the brane integrated over the internal space.
This stress is determined by the vacuum expectation value of the ${}_{D}^{D}$%
-component of the bulk energy-momentum tensor: $P=-\int d^{D_{2}}X\,\sqrt{%
\gamma }\langle 0|T_{D}^{\text{\textrm{(v)}}D}|0\rangle $. We expect that in
the presence of the surface energy the energy balance equation will be in
the form
\begin{equation}
dE=-PdV+\sum_{j=a,b}E_{j}^{\text{\textrm{(s)}}}dS^{(j)},\;E_{j}^{\text{%
\textrm{(s)}}}=\int d^{D_{2}}X\,\sqrt{\gamma }\varepsilon _{j}^{\text{%
\textrm{(s)}}},  \label{enbalance}
\end{equation}%
where $V$ is the $(D+1)$-volume in the bulk and $S^{(j)}$ is the $D$-volume
on the brane $y=j$ per unit coordinate volume in the $D_{1}$-dimensional
subspace,
\begin{eqnarray}
V &=&\int_{a}^{b}dye^{-Dk_{D}y}\int d^{D_{2}}X\,\sqrt{\gamma },  \label{Vj}
\\
S^{(j)} &=&e^{-Dk_{D}j}\int d^{D_{2}}X\,\sqrt{\gamma },\quad j=a,b.
\label{Sj}
\end{eqnarray}%
Combining equations (\ref{enbalance})-(\ref{Sj}), one obtains
\begin{equation}
\frac{\partial E}{\partial z_{j}}=\frac{n^{(j)}P^{(j)}-Dk_{D}E_{j}^{\text{%
\textrm{(s)}}}}{(k_{D}z_{j})^{D+1}},  \label{dEdzj}
\end{equation}%
with $P^{(j)}$ being the perpendicular vacuum stress on the brane at $y=j$.
The vacuum stresses normal to the branes can be presented in the form $%
p^{(j)}=p_{1}^{(j)}+p_{\text{\textrm{(int)}}}^{(j)}$, $j=a,b$, where $%
p_{1}^{(j)}$\ is the stress for a single brane at $y=j$ when the second
brane is absent and $p_{\text{\textrm{(int)}}}^{(j)}$ is induced by the
presence of the second brane. The latter determines the interaction forces
between the branes and is defined by the formula given in Ref. \cite{Saha06b}%
. Here we consider the corresponding quantity integrated over the internal
space:
\begin{eqnarray}
P_{\text{\textrm{(int)}}}^{(j)} &=&\int d^{D_{2}}X\,\sqrt{\gamma }p_{\text{%
\textrm{(int)}}}^{(j)}=k_{D}^{D+1}z_{j}^{D}\beta _{D_{1}}\sum_{\beta
}\int_{\lambda _{\beta }}^{\infty }du\,u  \notag \\
&&\times (u^{2}-\lambda _{\beta }^{2})^{D_{1}/2-1}\Omega _{j\nu
}(uz_{a},uz_{b})f_{\beta }^{(j)}(uz_{j}),  \label{pintj}
\end{eqnarray}%
with the notation
\begin{equation}
f_{\beta }^{(j)}(u)=\left( u^{2}-\nu ^{2}+2\frac{m^{2}}{k_{D}^{2}}\right)
B_{j}^{2}-D(4\zeta -1)A_{j}B_{j}-A_{j}^{2}.  \label{Fbetj}
\end{equation}%
As in the case of the induced surface energy density, the formula for the
interaction force can also be presented in the form%
\begin{eqnarray}
P_{\text{\textrm{(int)}}}^{(j)} &=&n^{(j)}(k_{D}z_{j})^{D+1}\beta
_{D_{1}}\sum_{\beta }\int_{\lambda _{\beta }}^{\infty }du\,u  \notag \\
&&\times (u^{2}-\lambda _{\beta }^{2})^{D_{1}/2-1}\left[ 1+\frac{%
2DB_{j}^{2}C_{j}}{B_{j}^{2}\left( x^{2}z_{j}^{2}+\nu ^{2}\right) -A_{j}^{2}}%
\right]  \notag \\
&&\times \frac{\partial }{\partial z_{j}}\ln \left\vert 1-\frac{\bar{I}_{\nu
}^{(a)}(uz_{a})\bar{K}_{\nu }^{(b)}(uz_{b})}{\bar{K}_{\nu }^{(a)}(uz_{a})%
\bar{I}_{\nu }^{(b)}(uz_{b})}\right\vert .  \label{Pjint}
\end{eqnarray}%
Now by taking into account expressions (\ref{Deltepsjnew}), (\ref{DeltaEhigh}%
), (\ref{Pjint}), it can be explicitly checked that the interference parts
in the vacuum energies and effective pressures on the branes obey the energy
balance equation
\begin{equation}
\frac{\partial \Delta E}{\partial z_{j}}=\frac{n^{(j)}P_{\text{\textrm{(int)}%
}}^{(j)}-Dk_{D}\Delta E_{j}^{\text{\textrm{(s)}}}}{(k_{D}z_{j})^{D+1}}.
\label{dDeltaEdzj}
\end{equation}%
The second term in the nominator on the right of this formula corresponds to
the additional pressure acting on the brane due to the nonzero external
curvature.

\section{Conclusion}

\label{sec:Conc}

Continuing our previous work \cite{Saha06a,Saha06b}, in the present paper we
have investigated the expectation value of the surface energy-momentum
tensor induced by the vacuum fluctuations of a bulk scalar field with an
arbitrary curvature coupling parameter satisfying Robin boundary conditions
on two parallel branes in background spacetime $\mathrm{AdS}_{D_{1}+1}\times
\Sigma $ with a warped internal space $\Sigma $. Vacuum stresses on the
brane are the same for both subspaces and the energy-momentum tensor on the
brane corresponds to the source of the cosmological constant type in the
brane universe. It is remarkable that the latter property is valid also for
the more general model with the mertic $g_{\mu \sigma }$ instead of $\eta
_{\mu \sigma }$ in line element (\ref{metric}). As an regularization
procedure for the surface energy density we employ the zeta function
technique. By using the residue theorem we have constructed an integral
representations for the partial zeta function corresponding to a given KK
mode along the internal subspace. This function is presented as the sum of
single brane and second brane induced parts [see formula (\ref{zet12})]. The
latter is finite at the physical point and the further analytical
continuation is necessary for the first term only. We have done this in
Section \ref{sec:1brane}. As the first step we subtract and add to the
integrand the leading terms of the corresponding asymptotic expansion for
large values of the argument and explicitly integrate the asymptotic part.
Further, for the regularization of the sum over the modes along the internal
space we use the local zeta function related to these modes. By making use
of the formula for the pole structure of this function, we have presented
the energy density on a single brane as the sum of the pole and finite
parts.These separate parts are determined by formulas (\ref{phi2Lpf1}), (\ref%
{phi2Lpf2}). The pole parts in the surface energy density are absorbed by
adding to the brane action the counterterms having the structure given by
Eq. (\ref{Counterterms}). The renormalized energy density on the
corresponding surface of a single brane is determined by formula (\ref%
{epsjren}), where the second term on the right presents the finite
renormalization part. The coefficients in this part cannot be determined
within the model under consideration and their values should be fixed by
additional renormalization conditions which relate them to observables.

Unlike to the single brane part, the surface energy density induced by the
presence of the second brane contains no renormalization ambiguities and is
investigated in Section \ref{sec:2brane}. This part is given by formula (\ref%
{emt2pl3}), or equivalently by formula (\ref{Deltepsjnew}). We have
investigated the induced energy density in various asymptotic regions for
the parameters of the model. In the limit $k_{D}\rightarrow 0$ we have
obtained the result for the geometry of two boundaries in the bulk $%
R^{(D_{1}-1,1)}\times \Sigma $. For the modes along $\Sigma $ with large KK
masses, the induced energy density is exponentially small. In particular,
for sufficiently small length scales of the internal space this is the case
for all nonzero KK modes and the main contribution comes from the zero mode.
For small interbrane distances, to the leading order, the induced energy
density is given by formula (\ref{epsjsmalldist}) and the contribution of
the induced part dominates the single brane parts. When the right brane
tends to the AdS horizon, $z_{b}\rightarrow \infty $, the induced energy
density on the left brane vanishes as $e^{-2\lambda _{\beta
}z_{b}}/z_{b}^{D_{1}/2}$ for the nonzero KK mode and like $%
z_{b}^{-D_{1}-2\nu }$ for the zero mode. In the same limit the corresponding
energy density on the right brane behaves as $z_{b}^{D_{2}+D_{1}/2+1}e^{-2%
\lambda _{\beta }z_{b}}$ for the nonzero KK mode and like $z_{b}^{D_{2}-2\nu
}$ for the zero mode. In the limit when the left brane tends to the AdS
boundary the contribution of a given KK mode into the induced energy density
vanishes as $z_{a}^{D+2\nu }$ and as $z_{a}^{2\nu }$ for the left and right
branes, respectively. For small values of the AdS curvature radius
corresponding to strong gravitational fields, for nonzero KK modes under the
conditions $\lambda _{\beta }z_{a}\gg 1$ and $\lambda _{\beta
}(z_{b}-z_{a})\gg 1$, the contribution to the induced energy density is
suppressed by the factor $e^{-2\lambda _{\beta }(z_{b}-z_{a})}$. For the
zero KK mode, the corresponding energy density integrated over the internal
space behaves as $k_{D}^{D_{1}+1}\exp [(D_{1}\delta _{j}^{a}+2\nu
)k_{D}(a-b)]$ for the brane at $y=j$ and is exponentially small. In the
model without the internal space we have similar behavior with $\nu $
replaced by $\nu _{1}$ and for a scalar field with $\zeta <\zeta
_{D+D_{1}+1} $ the suppression is relatively weaker.

In the model under discussion the hierarchy between the fundamental Planck
scale and the effective Planck scale in the brane universe is generated by
the combination of redshift and large volume effects. The corresponding
efective Newton's constant on the brane at $y=j$ is related to the
higher-dimensional fundamental Newton's constant by formula (\ref{GDj}) (see
also Eq. (\ref{Planckhierarchy}) for the ratio of the corresponding Planck
masses)\ and for large interbrane separations is exponentially small on the
brane $y=b$. We show that this mechanism also allows obtaining a naturally
small cosmological constant generated by the vacuum quantum fluctuations of
a bulk scalar. For large interbrane distances the ratio of the induced
cosmological constant to the the corresponding Planck scale quantity in the
brane universe is estimated by formula (\ref{LambDjest1}) \ and is
exponentially small. For the visible brane in the higher dimensional
generalization of the Randall-Sundrum two-brane model, this ratio is given
in terms of the effective and fundamental Planck masses by Eq.~(\ref{Lambonb}%
). We have considered two classes of models with the compactification scale
on the visible brane close to the fundamental Planck scale. For the first
one the higher dimensional Planck mass and the AdS inverse radius are of the
same order and in the second one a separation between these scales is
assumed. In both cases the corresponding interbrane distances generating the
hierarchy between the electroweak and Planck scales are smaller than those
for the model without an internal space and the required suppression of the
cosmological constant is obtained without fine tuning.

In Section \ref{sec:enbal} we have considered the total vacuum energy in the
region between the branes, evaluated as the sum of zero-point energies for
elementary oscillators. We have shown that this energy is equal to the sum
of surface and volume energies. The latter is evaluated as the integral of
the bulk energy density. Similar to the case of the surface energy, for the
evaluation of the total vacuum energy we have used the zeta function
regularization technique. This energy is presented as the sum of the single
brane and interference parts. We have extracted the pole parts of the single
brane vacuum energies and have discussed the corresponding renormalization
procedure. The divergences can be absorbed by adding the counterterms to the
brane action which have the structure similar to that for the surface
energy. The interference part in the vacuum energy is given by formula (\ref%
{DeltaEhigh}) and is regular for all nonzero interbrane distances. We have
investigated the asymptotic behavior of this part in various limiting cases.
For sufficiently small interbrane distances the total vacuum energy is
dominated by the interference part given by formula (\ref{DeltEsmalldist}).
In particular, from this formula it follows that in the case $\sigma =1$ any
minimum of the vacuum energy is local and the fixation of the interbrane
distance by the corresponding effective potential is stable only locally.
Further, we have shown that the induced vacuum densities and vacuum
effective pressures on the branes satisfy the standart energy balance
equation (\ref{enbalance}) with the inclusion of the surface terms.

\section*{Acknowledgments}

The work was supported by the Armenian Ministry of Education and Science
Grant No. 0124 and by PVE/CAPES Program.

\end{document}